\documentclass[12pt, aps, superscriptaddress, onecolumn, tightenlines]{revtex4-2}
\setcitestyle{super}

\usepackage[utf8]{inputenc}
\usepackage{graphicx}
\graphicspath{ {./figures/} }
\usepackage{textcomp,gensymb}
\usepackage{lineno}

\usepackage{prettyref}
\newrefformat{fig}{Fig.~\ref{#1}}
\newrefformat{exfig}{Extended Data Fig.~\ref{#1}}

\usepackage[normalem]{ulem}
\usepackage{color}
\usepackage{units}
\usepackage{times}
\usepackage{amsmath,amssymb}
\usepackage{comment}
\usepackage{hyperref}
\hypersetup{
    colorlinks=true,
    linkcolor=blue,
    filecolor=magenta,      
    urlcolor=cyan,
    citecolor=red,
    pdftitle={Spectroscopic Signatures of Strong Correlations and Unconventional Superconductivity in Twisted Trilayer Graphene},
    pdfpagemode=FullScreen,
}
\usepackage{wasysym}

\definecolor{orange}{rgb}{1,0.5,0}

\usepackage{amsmath,amssymb}

\renewcommand{\figurename}{\textbf{Fig.}}
\newcommand{\beginsupplement}{%
        \setcounter{table}{0}
        \renewcommand{\table}{\arabic{table}|}%
        \renewcommand{\thesection}{\arabic{section}}
        \setcounter{figure}{0}
        \renewcommand{\figurename}{\textbf{Extended Data Fig.}}
     }

\newcommand{\cCom}{\mathbin{\raisebox{0.5ex}{,}}}
\newcommand{\vk}{{\boldsymbol{k}}}
\newcommand{\vq}{{\boldsymbol{q}}}
\renewcommand{\vr}{{\boldsymbol{r}}}

\renewcommand{\vec}[1]{{\boldsymbol #1}} %
\newcommand{\caltechPH}{Department of Physics, California Institute of Technology, Pasadena, California 91125, USA}
\newcommand{\caltechAPH}{T. J. Watson Laboratory of Applied Physics, California Institute of Technology,
  1200 East California Boulevard, Pasadena, California 91125, USA}
\newcommand{\caltechTH}{Walter Burke Institute for Theoretical Physics, California Institute of Technology, Pasadena, California 91125, USA}
\newcommand{\caltechIQIM}{Institute for Quantum Information and Matter, California Institute of Technology, Pasadena, California 91125, USA}
\newcommand{\nims}{National Institute for Materials Science, Namiki 1-1, Tsukuba, Ibaraki 305 0044, Japan}

\begin{document}

\title{Spectroscopic Signatures of Strong Correlations and Unconventional Superconductivity in Twisted Trilayer Graphene}

\author{Hyunjin Kim}
\thanks{These authors contributed equally to this work.}
\affiliation{\caltechAPH}
\affiliation{\caltechIQIM}
\affiliation{\caltechPH}
\author{Youngjoon Choi}
\thanks{These authors contributed equally to this work.}
\affiliation{\caltechAPH}
\affiliation{\caltechIQIM}
\affiliation{\caltechPH}
\author{Cyprian Lewandowski}
\affiliation{\caltechIQIM}
\affiliation{\caltechPH}
\affiliation{\caltechTH}
\author{Alex Thomson}
\affiliation{\caltechIQIM}
\affiliation{\caltechPH}
\affiliation{\caltechTH}
\affiliation{Department of Physics, University of California, Davis, California 95616, USA}
\author{Yiran Zhang}
\affiliation{\caltechAPH}
\affiliation{\caltechIQIM}
\affiliation{\caltechPH}
\author{Robert Polski}
\affiliation{\caltechAPH}
\affiliation{\caltechIQIM}
\author{Kenji Watanabe}
\affiliation{\nims}
\author{Takashi Taniguchi}
\affiliation{\nims}
\author{Jason Alicea}
\affiliation{\caltechIQIM}
\affiliation{\caltechPH}
\affiliation{\caltechTH}
\author{Stevan Nadj-Perge}
\email{Correspondence: s.nadj-perge@caltech.edu}
\affiliation{\caltechAPH}
\affiliation{\caltechIQIM}

\begin{abstract}
\end{abstract}
\maketitle

  \textbf{ Magic-angle twisted trilayer graphene (MATTG) has emerged as a novel
    moir\'e material that exhibits both strong electronic correlations and
    unconventional superconductivity\cite{parkTunableStronglyCoupled2021,
      haoElectricFieldTunable2021}. However, spectroscopic studies of its
    electronic properties are lacking, and the nature of
    superconductivity and the corresponding order parameter in this system 
    remain elusive. Here we perform high-resolution scanning
    tunneling microscopy and spectroscopy of MATTG and reveal extensive regions
    of atomic reconstruction that favor mirror-symmetric stacking. In these
    regions we observe a cascade of symmetry-breaking electronic transitions and
    doping-dependent band structure deformations similar to those realized in
    magic-angle bilayers, as expected theoretically given the commonality of
    flat bands\cite{khalafMagicAngleHierarchy2019,
      liElectronicStructureSingleTwist2019}. More strikingly, in a density
    window spanning two to three holes per moir\'e unit cell, spectroscopic
    signatures of superconductivity are manifest as pronounced dips in the
    tunneling conductance at the Fermi level accompanied by coherence peaks that
    become gradually suppressed at elevated temperatures and magnetic fields.
    The observed evolution of the conductance with doping is consistent with a
    gate-tunable transition from a gapped to a nodal superconductor, which we
    show theoretically is compatible with a sharp transition from a
    Bardeen-Cooper-Schrieffer (BCS) to a Bose-Einstein-condensation (BEC)
    superconductor with a nodal order
    parameter. Within this doping window we
    also detect peak-dip-hump structures suggesting that superconductivity is
    driven by strong coupling to bosonic modes of MATTG. Our results pave the
    way for further understanding of superconductivity and correlated states in
    graphene-based moir\'e structures beyond twisted bilayers, where unconventional 
    superconductivity and nodal pairing are also reported\cite{yazdaniSpectroscopicSignaturesNodal2021}.}

\begin{figure}[p]
\begin{center}
    \includegraphics[width=16cm]{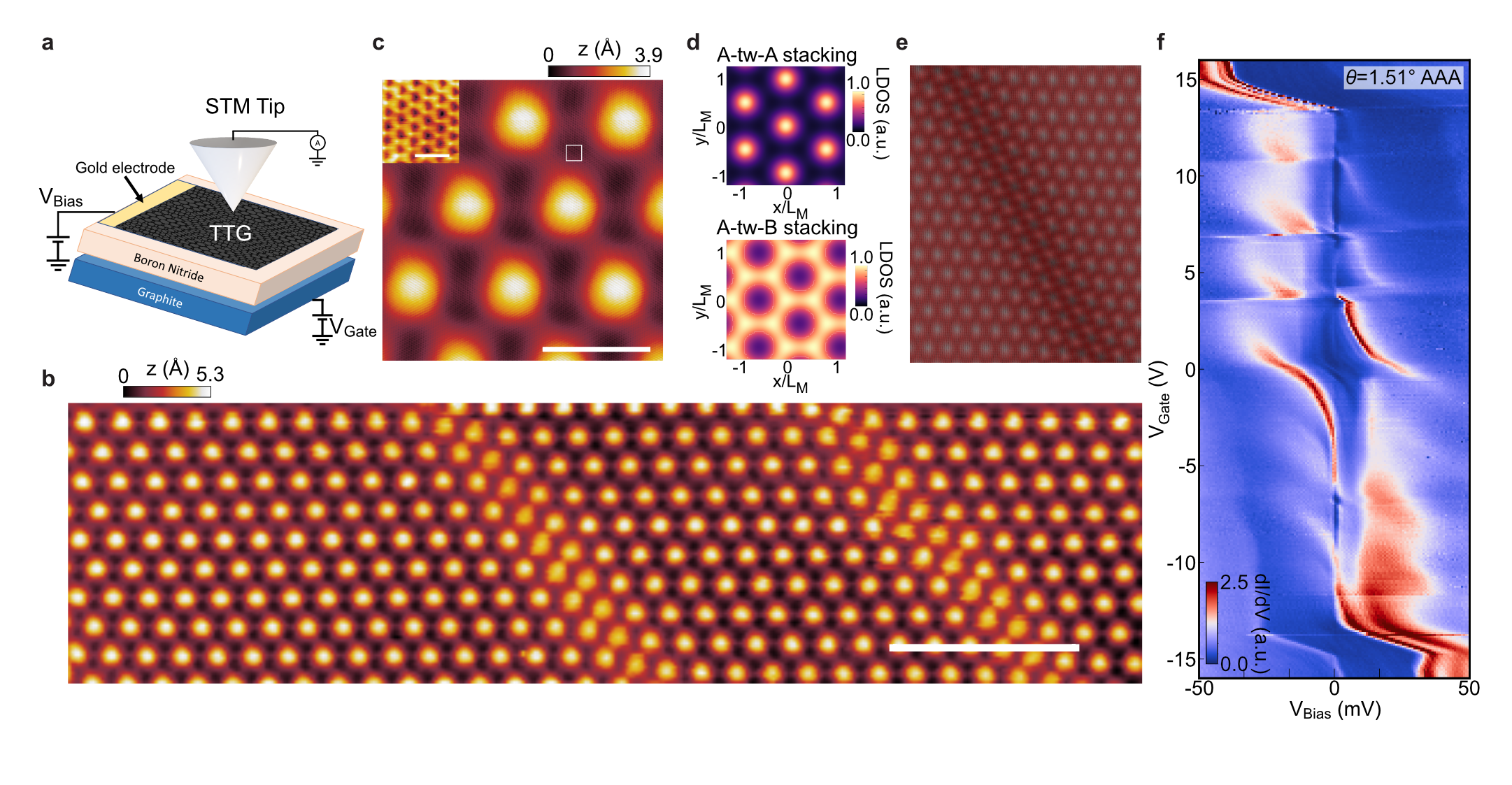}
\end{center}
\caption{ {\bf Topography and spectroscopy of MATTG at zero magnetic field.}
  {\bf a}, Schematics of the STM experiment. MATTG is placed on an hexagonal
  Boron Nitride (hBN) substrate and doping is controlled by a graphite back
  gate. {\bf b}, $290$ nm by $80$ nm area where two stripes separated by
  approximately $100$ nm are observed (tunneling set point parameters:
  $V_\mathrm{{Bias}} = 100$~mV, $I = 20$~pA; scale bar 50 nm). {\bf c}, $26$ nm
  by $26$ nm topography showing moir\' e lattices with corresponding moire
  length of approximately $9$ nm (scale bar $10$ nm). The inset shows the
  atomic-scale hexagonal lattice of carbon atoms (scale bar $0.5$ nm). {\bf d},
  Calculated local density of states (LDOS) at charge neutrality originating
  from the bands within approximately $\pm 50$~meV energy window for A-tw-A
  (upper panel) and A-tw-B (lower panel) stacking. While in principle various
  configurations could arise, the A-tw-A stacking, where first and third layers
  are aligned, is seen experimentally. The peaks in LDOS correspond to AAA
  stacked regions where carbon atoms from three graphene layers are aligned.
  {\bf e}, Simulated atomic distribution of MATTG with the first and third
  layers strained with respect to each other (See Methods,
  section~\ref{methods: stripe} for simulation details). {\bf f}, Tunneling
  conductance ($dI/dV$) spectroscopy as a function of $V_{\mathrm{Gate}}$ at twist angle $\theta = 1.51\degree$ on an AAA site at
  $T = 400$~mK. Clear signatures of symmetry breaking cascades, similar to
  twisted gaphene bilayers\cite{wongCascadeElectronicTransitions2020,
    choiCorrelationdrivenTopologicalPhases2021}, are observed.}
\label{fig: fig1}
\end{figure}

  Figure \ref{fig: fig1}a,b,c shows a schematic of the scanning tunneling
  microscopy (STM) setup and MATTG topography formed by alternatingly rotating
  three graphene layers by
  $\theta=1.5\degree$\cite{khalafMagicAngleHierarchy2019,
    parkTunableStronglyCoupled2021, haoElectricFieldTunable2021}, resulting in a
  moir\'e wavelength of $L_{m} = a/[2\sin(\theta/2)]\approx9$ nm, where
  $a=0.246$~nm is the graphene crystal lattice (see Methods, sections
  \ref{methods: fabrication} and \ref{methods: stm} for fabrication and
  measurement details). Since MATTG is composed of three layers, two independent
  moir\'e patterns can in principle arise and, moreover, possible offsets
  between the first and third layers, could result in even more complex
  outcomes. Surprisingly, however, we consistently observe a unique triangular
  moir\'e lattice, with no sign of an additional underlying moir\'e pattern,
  signaling the formation of a single predominantly A-tw-A configuration in
  which the first and third layers are aligned and second layer is twisted by
  $\theta$ (\prettyref{fig: fig1}c,d). This observation suggests that mirror
  symmetric A-tw-A stacking is preferred, in line with previous ab-initio theory
  calculations\cite{carrUltraheavyUltrarelativisticDirac2020} and transport
  measurements\cite{parkTunableStronglyCoupled2021,
    haoElectricFieldTunable2021}. Additionally, in large-scale topographies, we
  occasionally observe stripe-like features (\prettyref{fig: fig1}b) that are
  not reported in twisted bilayers. We attribute these stripes to domain
  boundaries where strain in the top and bottom layers arises as a result of the
  the atomic reconstruction necessary to maintain A-tw-A stacking across the
  domains (\prettyref{fig: fig1}e; Methods, section \ref{methods: stripe}).

  Spectroscopy of MATTG (\prettyref{fig: fig1}f) upon electrostatic doping
  (controlled by the gate voltage $V_{\rm Gate}$) is similar to magic-angle
  twisted bilayer graphene (MATBG) in many respects---a reflection of the
  alternating-angle stacking of the trilayer, which conspires to form
  spin/valley-degenerate flat bands, together with additional dispersive Dirac
  cones\cite{khalafMagicAngleHierarchy2019,
    carrUltraheavyUltrarelativisticDirac2020}. The two Van Hove singularities
  (VHSs) originating from those flat bands, detected as peaks in tunneling
  conductance $dI/dV$, are pushed apart at the charge neutrality point (CNP,
  $\nu = 0$) compared to full filling of four electrons (holes) per moir\'e unit
  cell ($\nu = \pm 4$). The approximately fivefold change in VHS separation
  indicates that the partially filled flat band structure is largely determined
  by electronic correlations in analogy with the behaviour seen in
  MATBG\cite{kerelskyMaximizedElectronInteractions2019,choiElectronicCorrelationsTwisted2019,
    xieSpectroscopicSignaturesManybody2019, jiangChargeOrderBroken2019}. A
  well-developed cascade of flavor symmetry-breaking phase
  transitions\cite{zondinerCascadePhaseTransitions2020,
    wongCascadeElectronicTransitions2020} is also observed (\prettyref{fig:
    fig1}f). The overall spectroscopic similarities between MATTG and MATBG
  suggest that the flat bands in MATTG dominate the local density of states
  (LDOS) in this regime. We do nevertheless detect subtle signatures of the
  expected additional Dirac cones. Most obviously, contrary to twisted bilayers,
  at $\nu=\pm4$ the LDOS is neither completely suppressed nor accompanied by
  quantum dot formation\cite{choiCorrelationdrivenTopologicalPhases2021} (see
  \prettyref{exfig: fig1})---indicating the presence of gapless states
  intervening between the flat bands and remote dispersive bands.

\begin{figure}[p]
\begin{center}
    \includegraphics[width=15cm]{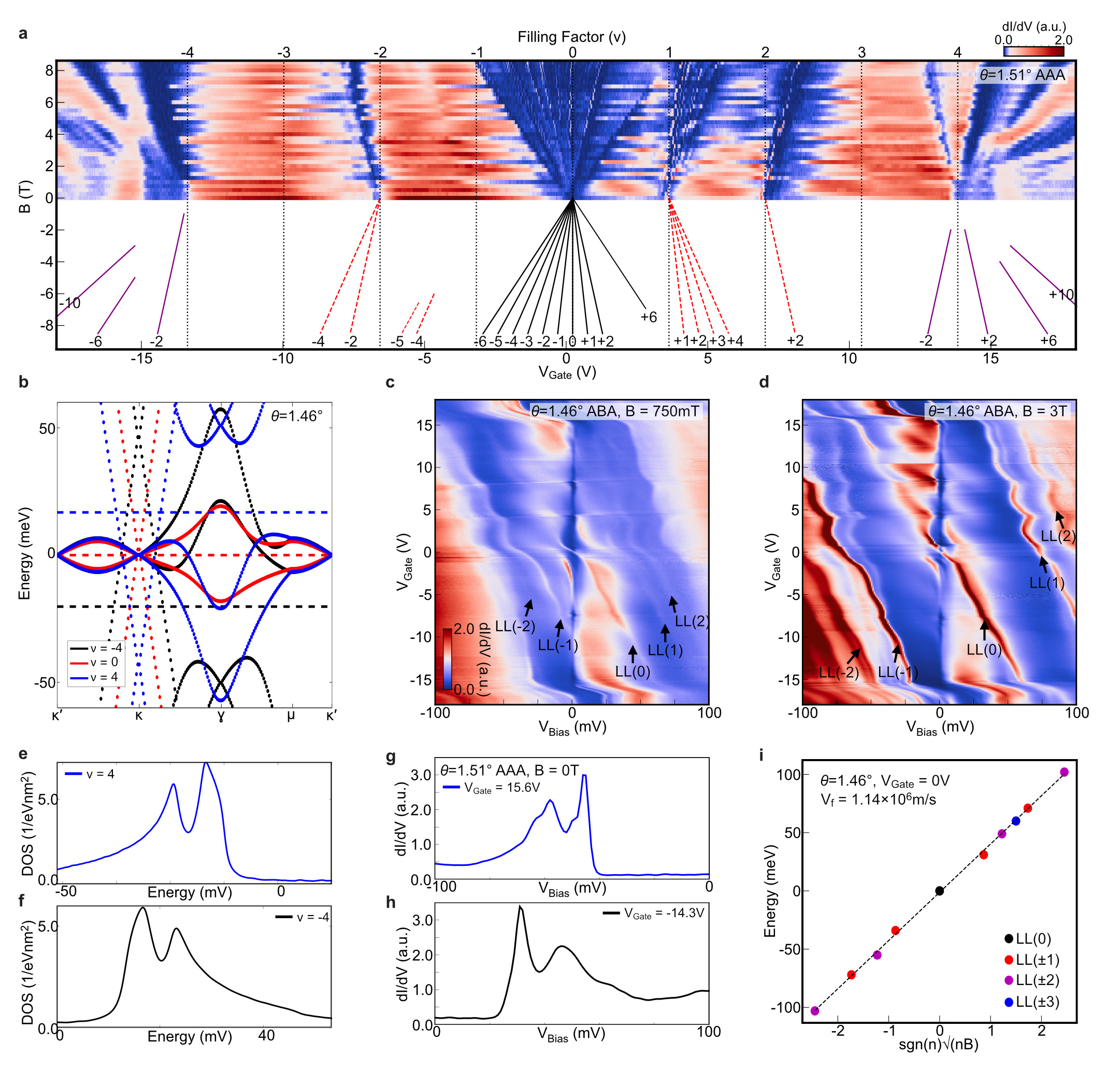}
\end{center}
\caption{{\bf LDOS Landau fan diagram and doping-dependent band deformations in
    MATTG.} {\bf a}, LDOS Landau fan
  diagram\cite{choiCorrelationdrivenTopologicalPhases2021} measured on an AAA
  site. The negative magnetic field fan shows the corresponding schematic of
  gaps between LLs emanating from the CNP (black); gaps emanating from non-zero
  integer fillings (red); and gaps between LLs from the dispersive bands
  (purple). {\bf b}, Calculated MATTG band structure taking into account Hartree
  corrections. Horizontal dashed lines represent the positions of the Fermi levels at each doping. Electron (hole) doping shifts the Dirac-like band towards
  negative (positive) energy relative to the flat band (see also Methods,
  section \ref{methods: Hartree}). {\bf c, d}, Point spectroscopy on an ABA site (in between AAA sites)
  at finite magnetic fields $B = 0.75~\text{T}$ ({\bf c}) and $B = 3~\text{T}$
  ({\bf d}). Black arrows indicate LLs identified to originate
  from the additional Dirac cones characteristic of MATTG. {\bf e}, {\bf f},
  Calculated density of states with Hartree corrections at $\nu = 4$ ({\bf e})
  and $\nu = -4$ ({\bf f}) for $\theta = 1.51\degree$ at $B = 0~\text{T}$. {\bf
    g}, {\bf h}, Point spectra taken at an AAA site at $B = 0~\text{T}$ near
  $\nu = 4$ ($V_{\rm Gate} = 15.6~\text{V}$, {\bf g}) and $\nu = -4$
  ($V_{\mathrm{Gate}} = -14.3~\text{V}$, {\bf h}). Note the asymmetric profile as expected from ({\bf e, f}). {\bf i}, Energies of LLs
  extracted from ({\bf c, d}) at $V_{\mathrm{Gate}}=0$~V and plotted versus
  ${\rm sgn}(n) \sqrt{\vert n \vert B}$, with $n$ is the LL index, showing agreement with
  expectations from a Dirac dispersion. All data in this figure are taken within a
  $100\times100$~nm$^2$ MATTG area with average $\theta=1.48\pm0.03\degree$. The
  angles shown in the panels are obtained from measuring the exact distances
  between the closest AAA sites. Measurements are taken at $T = 2~\text{K}$. }
\label{fig: fig2}
\end{figure}

  The LDOS at the Fermi level measured at finite magnetic
  fields\cite{choiCorrelationdrivenTopologicalPhases2021} provides further
  signatures of the additional Dirac cones in MATTG (\prettyref{fig: fig2}a). We
  resolve clear Landau fans emanating from zero field around $\nu = 0, \pm 4$
  along with $\nu = +1, \pm 2$; the latter signal Fermi surface reconstructions
  due to flavour symmetry-breaking transitions in agreement with conclusions of
  transport studies\cite{parkTunableStronglyCoupled2021,
    haoElectricFieldTunable2021}. The main fan sequence originating from
  $\nu = +4$ is $+2, +6, \dots$ ($-2, -6, \dots$ for $\nu = -4$) instead of the
  $0, +4, \dots$ pattern typically seen in MATBG devices. The relative
  Chern-number shift of $2$ naturally arises from the zeroth Landau level (LL)
  associated with the additional Dirac cones, which contribute to the total
  Chern number at $\nu = \pm 4$. Finite-bias spectroscopy in magnetic fields
  more directly exposes the presence of additional Dirac cones in the spectrum
  (\prettyref{fig: fig2}c,d). Here we can clearly identify the
  $N=0,\pm1,\pm2,\dots$ Landau levels originating from the Dirac dispersion; the
  increase of Landau level separation with field (\prettyref{fig: fig2}f)
  confirms the linear dispersion and yields a monolayer-graphene Dirac velocity
  in agreement with theoretical
  expectations\cite{liElectronicStructureSingleTwist2019,
    carrUltraheavyUltrarelativisticDirac2020}.

  Spectroscopy at finite magnetic fields additionally uncovers filling-dependent
  band structure renormalization in
  MATTG\cite{fischerUnconventionalSuperconductivityMagicAngle2021,
    phongBandStructureSuperconductivity2021}. The effect originates from the
  inhomogeneous real-space charge distribution associated with different energy
  eigenstates: the majority of the weight of the flat-band states (including
  those near the VHS) are spatially located on the AAA moir\'e sites, whereas
  the additional Dirac cones and flat-band states in the immediate vicinity of
  the $\gamma$ point are more uniformly distributed (see \prettyref{exfig:
    fig2}). Electrostatic doping thereby gives rise to a Hartree potential that
  modifies the band structure in a manner that promotes charge uniformity
  throughout the unit cell. In twisted bilayer graphene it was
  found\cite{choiInteractiondrivenBandFlattening2021} that this potential
  generates additional band deformations
  \cite{guineaElectrostaticEffectsBand2018,rademakerChargeSmootheningBand2019,
    goodwinHartreeTheoryCalculations2020,
    calderonInteractions8orbitalModel2020}. Our simulations capture a similar
  band-renormalization in MATTG accompanied by a displacement of the additional
  Dirac cones away from the flat
  bands\cite{fischerUnconventionalSuperconductivityMagicAngle2021,
    phongBandStructureSuperconductivity2021} (\prettyref{fig: fig2}b). Both
  effects---band deformations (\prettyref{fig: fig2}e-h) and the relative Dirac
  cone shift---are clearly confirmed in our measurements. Importantly, the
  position of the Dirac point obtained from tracking the zeroth Landau level
  (\prettyref{fig: fig2}c,d) falls within $\pm50$~meV depending on the exact
  doping; it resides below the lower flat-band VHS at $\nu = +4$ but moves above
  the upper flat-band VHS at $\nu = -4$. This pronounced shift may explain the
  large bandwidth estimate of $> 100$~meV from
  Ref.~\citenum{parkTunableStronglyCoupled2021} (see Methods, section
  \ref{methods: Hartree}B,C for additional discussion). Finally, we note that
  the Landau levels from the Dirac cones appear unaltered by the cascade of
  phase transitions in the flat bands, suggesting that the flat-band and Dirac
  sectors are not strongly coupled by
  interactions\cite{christosCorrelatedInsulatorsSemimetals2021}.

\begin{figure}[p]
\begin{center}
    \includegraphics[width=15cm]{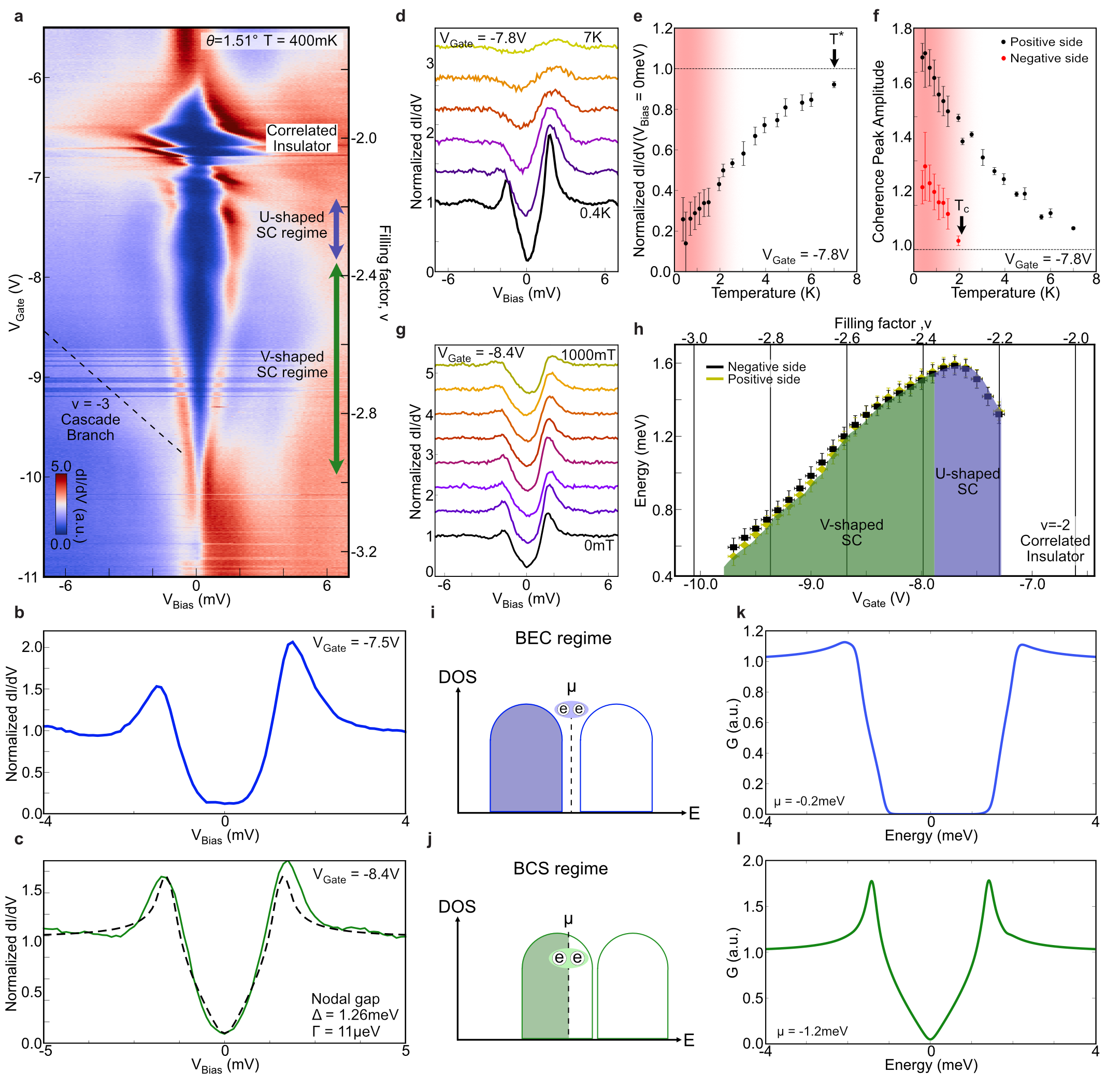}
\end{center}
\caption{{\bf Spectroscopic gap in the $\mathbf{-3<\nu<-2}$ range and signatures
    of unconventional superconductivity.} {\bf a}, Spectra near an AAA site
  (same area as \prettyref{fig: fig2}a). Purple and green arrows denote $\nu$
  range over which U- and V-shaped tunneling spectra, accompanied by clear
  coherence peaks, are observed. {\bf b}, {\bf c}, Normalized spectra showing U-shaped
  ({\bf b}) and V-shaped ({\bf c}) tunneling suppression. The data are
  normalized by a polynomial background, and fit to the Dynes formula ({\bf c})
  with a nodal superconducting order parameter (see Methods, section
  \ref{methods: fitting}). {\bf d}, Temperature dependence of the spectrum (lines correspond to $T=0.4, 2, 3, 4.5, 5.6, 7$~K). {\bf
    e}, Normalized zero-bias conductance vs.~temperature; $T^*$ indicates the
  temperature at which the zero-bias conductance reaches 90\% of the conductance
  outside the gap. {\bf f}, Coherence-peak amplitude vs.~temperature from
  normalized spectra on the electron (black) and hole (red) side. The hole-side
  coherence peak gets fully suppressed around $T_c\approx 2-2.5$~K. ({\bf d-f}) are from the same dataset as \prettyref{exfig: fig5}h-k. {\bf g},
  Magnetic-field dependence of the spectrum (lines
  correspond to $B = 0, 100, 200, 300, 400, 600, 800, 1000$~mT), from the same dataset as \prettyref{exfig: fig5}a-d. {\bf h},
  Gap size $\Delta$ vs.~$\nu$ ($V_{\rm Gate}$) extracted from ({\bf
    a}) separately for electron (yellow) and hole (black) side coherence peaks.
  Color coding of different regions matches ({\bf a}). {\bf i-l}, Proposed BEC-BCS transition ({\bf i}, {\bf j}) mechanism that
  qualitatively reproduces U- and V-shaped spectra ({\bf k}, {\bf l}); see main
  text and Methods, section \ref{methods:bec_bcs_transition}. }
\label{fig: fig3}
\end{figure}

  Having established the foundational properties of MATTG band structure, we now
  turn to the doping range $-3 \lesssim \nu \lesssim -2$, where significant
  suppression of the tunneling conductance is observed (\prettyref{fig: fig3}a).
  We identify two main doping regions---one at $-2.1<\nu<-1.9$ and the other at
  $-3<\nu<-2.2$. The former interval, around $\nu\approx-2$, exhibits a
  correlation-induced gap accompanied by Coulomb diamonds and nearly horizontal
  resonance peaks, signaling the formation of quantum dots and a correlated
  insulating state\cite{jungEvolutionMicroscopicLocalization2011,
    choiCorrelationdrivenTopologicalPhases2021}, despite the presence of the
  additional Dirac cones.
  
  Throughout the second interval, $-3<\nu<-2.2$, the tunneling conductance
  minimum is well-pinned to the Fermi energy ($V_{\rm Bias} = 0$) despite the
  large change in filling. Strikingly, this suppression is accompanied by peak
  structures symmetrically placed around the Fermi energy as line traces show in
  \prettyref{fig: fig3}b,c (note that the spectra taken at $-2.1 < \nu < -1.9$
  do not exhibit these symmetric peaks; see \prettyref{exfig: fig4}).
  The presence of such sharp narrow peaks---which strongly resemble coherence
  peaks in superconductors and occur in the filling range where transport
  experiments observe
  superconductivity\cite{parkTunableStronglyCoupled2021,haoElectricFieldTunable2021}---leads
  us to attribute this spectroscopic signature to superconductivity in MATTG.

  Temperature and magnetic field dependence of the tunneling spectra
  (\prettyref{fig: fig3}d-g) corroborates the connection to superconductivity
  while also establishing its unconventional nature. As the temperature is
  increased, the coherence peaks on both sides of the Fermi energy subside
  gradually until $2-2.5$~K (close to the maximum critical temperature reported
  in transport\cite{parkTunableStronglyCoupled2021}), where the hole-side peak
  completely disappears (\prettyref{fig: fig3}d,f) and the zero-bias conductance
  exhibits a visible upturn (\prettyref{fig: fig3}e; see also \prettyref{exfig:
    fig5} for more data). Suppressed zero-bias conductance together with a
  significantly broadened electron-side peak nevertheless survives at this
  temperature; both features are washed out only around $T^{*}\approx7$~K
  (\prettyref{fig: fig3}e,f). Persistent conductance suppression beyond the
  disappearance of coherence peaks is typically interpreted as evidence of a
  pseudogap phase characteristic of unconventional superconductors such as
  cuprates or thin films of disordered
  alloys\cite{eaglesPossiblePairingSuperconductivity1969,
    rennerPseudogapPrecursorSuperconducting1998} (see \prettyref{exfig: fig6}
  for data near $\nu=+2$). Our observation of two different temperature scales
  is consistent with the existence of superconducting and pseudogap phases in
  MATTG. In any case, the gradual disappearance of the coherence peak with
  temperature reaffirms its superconducting origin.

  Denoting the coherence peak-to-coherence peak distance as $2\Delta$, we find
  maximal $\Delta \approx 1.6~\text{meV}$ near $\nu=-2.4$ (\prettyref{fig:
    fig3}h). The overall doping dependence of the spectroscopic gap resembles
  the doping dependence of the critical temperature
  $T_C$\cite{parkTunableStronglyCoupled2021, haoElectricFieldTunable2021}, which
  also peaks around $\nu \approx -2.4$, suggesting a correlation between these
  two quantities. The maximal critical temperature $T_C \approx 2-2.5$~K from
  transport\cite{parkTunableStronglyCoupled2021} yields a ratio
  $2\Delta/k_{B}T_{C} \approx 15-19$ ($k_B$ is Boltzmann's constant) that far
  exceeds the conventional BCS value ($\approx 3.5$)---highlighting the
  strong-coupling nature of superconductivity in MATTG. The measured
  spectroscopic gaps also imply a maximum Pauli limit of $\sim10~\text{T}$ for
  the destruction of spin-singlet superconductivity.

  The coherence peak height at base temperature ($T=400$~mK) also gradually
  decreases with perpendicular magnetic field, similar to tunneling conductance
  measurements through MATBG
  junctions\cite{rodan-legrainHighlyTunableJunctions2021}. We observe that the
  coherence peaks are greatly diminished by $1$~T and therefore infer a critical
  field $B_{C} \gtrsim 1~\text{T}$ at $\nu \approx -2.4$ (\prettyref{fig:
    fig3}g; see also \prettyref{exfig: fig5}). This result is compatible with
  the small Ginzburg-Landau coherence length of
  $\xi_{\rm GL}\approx 12~\text{nm}$ reported around optimal
  doping\cite{parkTunableStronglyCoupled2021} upon using the naive estimate
  $B_C\approx \Phi_{0}/{2\pi \xi_{\rm GL}^{2}} \sim 2~\text{T}$, where $\Phi_0$
  is the magnetic flux quantum. Note that LDOS suppression without coherence
  peaks persists up to much larger fields (\prettyref{exfig: fig5}f,g).

  Interestingly, suppressed tunneling conductance within the coherence peaks
  typically evolves from a U-shaped profile at $-2.4 \lesssim \nu < -2.2$
  (\prettyref{fig: fig3}b) to a V-shaped profile at
  $-3 \lesssim \nu \lesssim -2.4$ (\prettyref{fig: fig3}c), suggesting two
  distinct superconducting regimes. Magnetic-field dependence of the tunneling
  conductance further distinguishes these regimes: the field more efficiently
  suppresses the spectroscopic gap in the V-shaped window compared to the
  U-shaped window (\prettyref{exfig: fig5}). The V-shaped tunneling spectra
  resemble that of cuprates and can be well-fit using the standard Dynes
  formula\cite{dynesDirectMeasurementQuasiparticleLifetime1978} with a pairing
  order parameter that yields gapless nodal excitations as reported in twisted bilayer graphene\cite{yazdaniSpectroscopicSignaturesNodal2021} (\prettyref{fig: fig3}c
  and \prettyref{exfig: fig7}; see Methods, section \ref{methods: fitting}). The
  enhanced conductance suppression of the U-shaped spectra instead suggests the
  onset of a fully gapped superconducting state. One logical possibility is that
  the U- and V-shaped regimes admit distinct superconducting order parameter
  symmetries that underlie a transition from a gapped to gapless paired state on
  hole doping (similar behavior has been proposed for cuprates
  \cite{yehEvidenceDopingDependentPairing2001}). We stress, however, that a
  standard isotropic s-wave pairing order parameter fails to adequately fit the
  U-shaped spectra, though reasonable agreement can be obtained by postulating a
  mixture of $s$-wave and nodal order parameters or a $d+id$-like order
  parameter (see Methods, section \ref{methods: fitting} and \prettyref{exfig:
    fig7}).

  We point here to an alternative explanation whereby the U- to V-shaped regimes
  can be understood in the context of BEC and BCS phases with a \emph{single}
  nodal order parameter. In this scenario, starting from the correlation-induced
  gapped flat bands at $\nu = -2$, hole doping initially introduces strongly
  bound Cooper pair `molecules,' rather than simply depleting the lower flat
  band; i.e., the chemical potential remains within the gap of the correlated
  insulator (\prettyref{fig: fig3}i). Condensing the Cooper pair molecules
  yields a BEC-like superconducting state that we assume exhibits a nodal order
  parameter. Crucially, the original correlation-induced flat-band gap
  nevertheless precludes gapless quasiparticle excitations. Further hole doping
  eventually begins depleting the lower flat band (\prettyref{fig: fig3}j), at
  which point the system transitions to a BCS-like superconductor. Here, Cooper
  pair formation onsets at the Fermi energy, and the nodal order parameter
  allows for gapless quasiparticle excitations. (When compared against a BEC
  phase, we use `BCS' to describe a superconductor for which the chemical
  potential intersects a band, independent of the pairing mechanism or coupling
  strength.) The gapped versus gapless distinction implies that the U- and
  V-shaped regimes are separated by a clear
  \emph{transition}\cite{botelhoLifshitzTransitionWave2005, borkowskiEvolutionBCSBoseEinstein2001} as opposed to the
  well-studied BEC-BCS \emph{crossover}\cite{chenBCSBECCrossover2005,
    randeriaCrossoverBardeenCooperSchriefferBoseEinstein2014} operative when
  both regimes are fully gapped and not topologically distinct.

  We phenomenologically model such a transition by considering the tunneling
  conductance of a system with electron and hole bands that experience
  doping-dependent band separation and nodal pairing chosen to mimic experiment;
  for details see Methods, section \ref{methods:bec_bcs_transition}. In the fully gapped
  BEC phase, this model yields U-shaped tunneling spectra (\prettyref{fig:
    fig3}k) that qualitatively match the measured conductance. Indeed, as in
  experiment, the conductance gap profile does not fit an isotropic $s$-wave
  pairing amplitude well due to the additional structure from the nodal order
  parameter. When the system enters the BCS phase (the chemical potential lies
  inside the band), the gapless nodal BCS phase instead yields a V-shaped
  tunneling profile (\prettyref{fig: fig3}l) that also qualitatively matches the
  experiment. This interpretation of the U- to V-shaped transition is bolstered
  by transport measurements\cite{parkTunableStronglyCoupled2021} that reveal two
  regimes for the Ginzburg-Landau coherence length (see Methods, section
  \ref{methods:bec_bcs_transition}).

\begin{figure}[p]
\begin{center}
    \includegraphics[width=16.5cm]{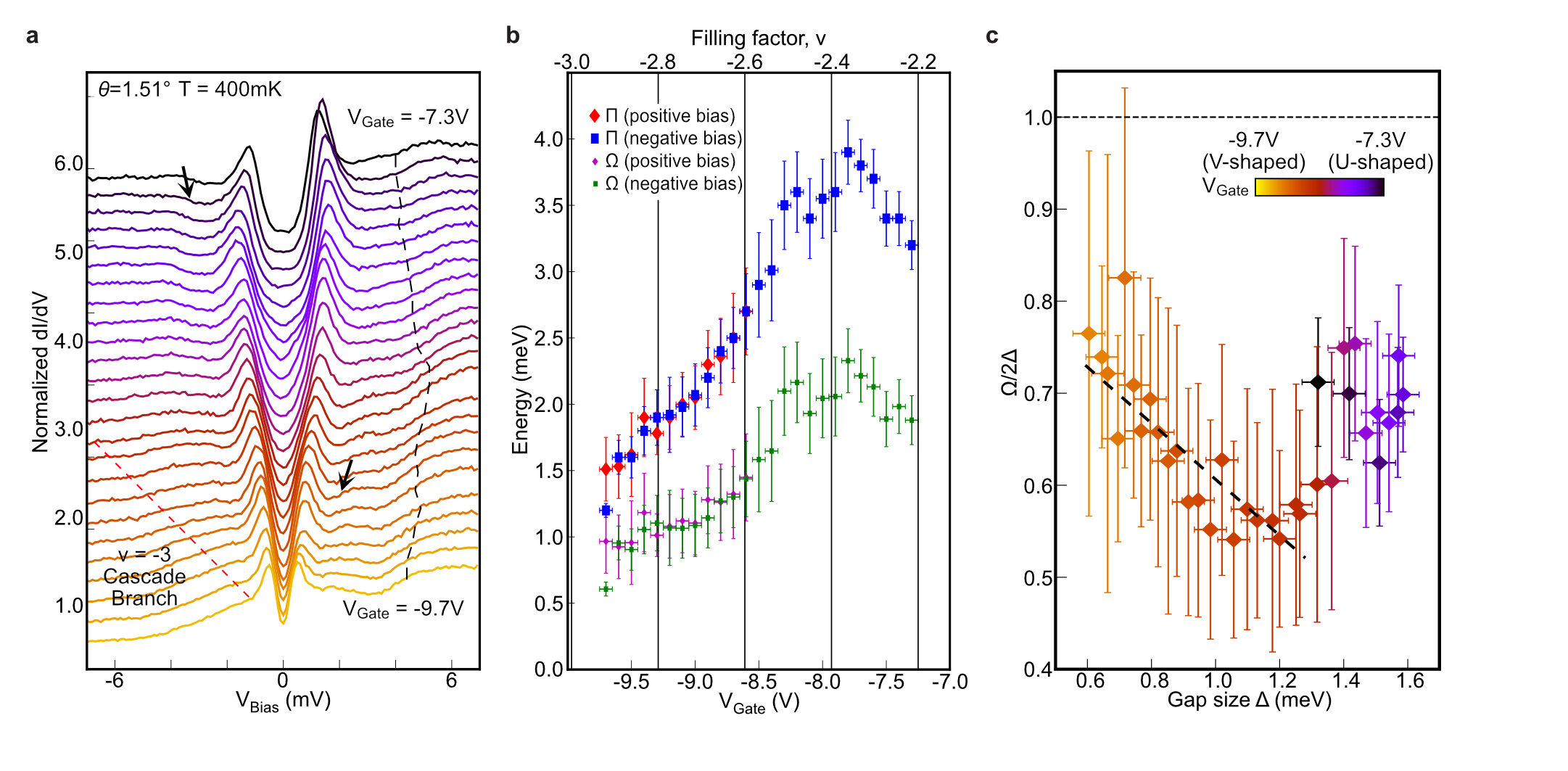}
\end{center}
\caption{{\bf Peak-dip-hump structure in MATTG}. {\bf a}, Line traces showing
  point spectra for $V_{\rm Gate}$ ranging from $-9.7$~V to $-7.3$~V (same
  dataset as \prettyref{fig: fig3}a). Each spectrum is divided by the mean
  value for clarity. Red dashed line indicates the LDOS peak originating
  from the sub-band that abruptly shifts due to the cascade near
  $\nu = -3$; black dashed line indicates the shoulder of the upper flat
  band VHS. Black arrows denote the position of hole-side and electron-side
  dip-hump structure identified from the local minimum/maximum in
  $d^2I/dV^2$. {\bf b},  Extracted energy $\Pi$ of the electron-side (red)
  and hole-side (blue) dip-hump structure and corresponding energy $\Omega$
  of the bosonic mode on the electron side (purple) and hole side (green)
  versus filling factor ($V_{\rm Gate})$. {\bf c}, Ratio $\Omega/2\Delta$
  plotted versus $\Delta$ for both electron- and hole-side bosonic
  excitations. The black dashed line is a linear regression of the data at
  $V_{\rm Gate}$ ranging from $-9.7~\text{V}$ to $-8.6~\text{V}$ that shows
  anticorrelation of the two quantities for fillings at which V-shaped
  tunneling spectra are observed.}
\label{fig: fig4}
\end{figure}

\begin{figure}[p]
\begin{center}
    \includegraphics[width=16cm]{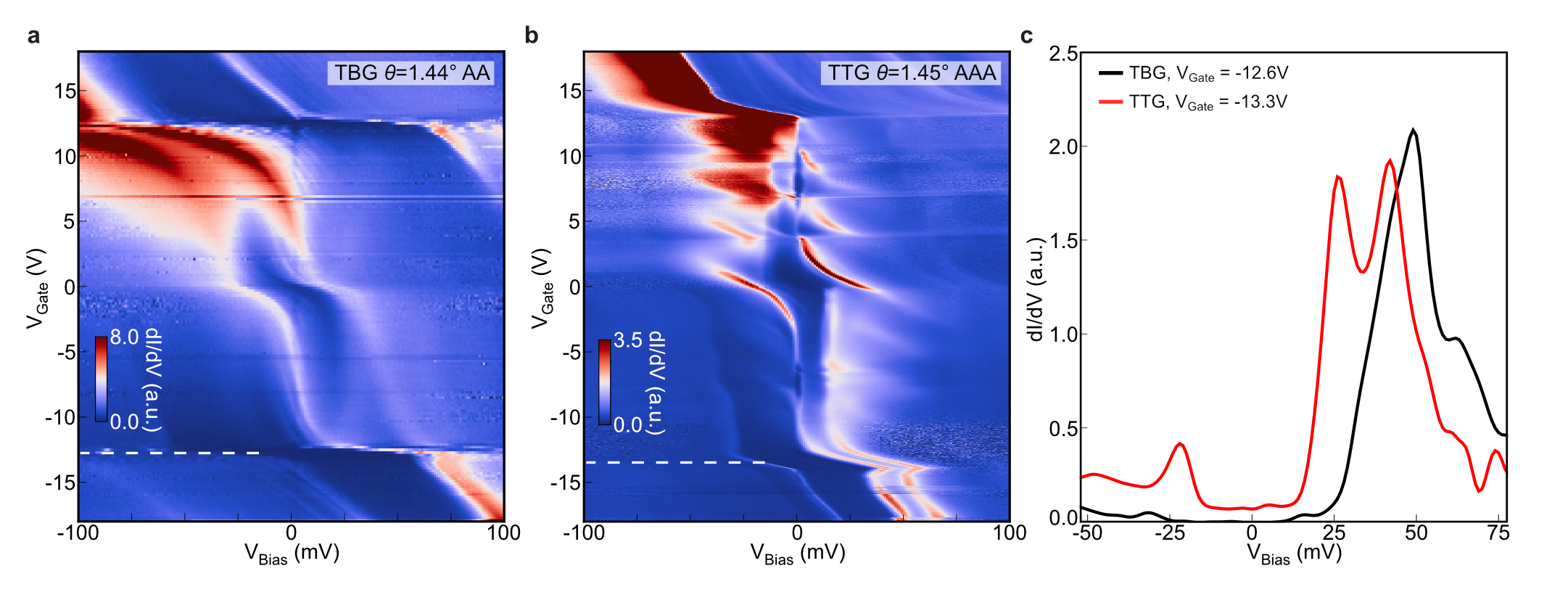}
\end{center}
\caption{{\bf Spectroscopy of twisted bilayer and twisted trilayer graphene.} 
{\bf a}, Point spectra of twisted bilayer graphene (TBG) on an AA site at a 
twist angle $\theta = 1.44\degree$, from a bilayer region found in 
the same sample. {\bf b}, Point spectra of twisted trilayer graphene (TTG) 
on an AAA site at a twist angle $\theta = 1.45\degree$. Unlike TBG 
at the similar angle, signatures of correlations, such as enhancement of VHS 
separations at charge neutrality and cascade of flavor symmetry breaking, 
are observed. {\bf c}, Linecuts taken from {\bf a} and {\bf b} around 
$\nu = -4$ (white dashed lines). While the $dI/dV \sim \text{LDOS}$ 
between the flat bands and the remote band is zero for TBG, the value is 
finite for TTG due to the existence of the additional Dirac cones.}
\label{exfig: fig1}
\end{figure}

  Adjacent to the coherence peaks, we observe dip-hump features in the tunneling
  conductance that persist over a broad doping range (\prettyref{fig: fig4}).
  The positive and negative voltage dips are typically symmetric in energy,
  independent of filling---ruling out the possibility that the dip-hump
  structure is intrinsic to background density of states. Similar dip-hump
  features are observed spectroscopically in a range of both conventional
  strongly coupled phonon
  superconductors\cite{schriefferEffectiveTunnelingDensity1963,mcmillanLeadPhononSpectrum1965}
  as well as unconventional cuprate, iron-based and heavy fermion
  superconductors\cite{leeInterplayElectronLattice2006,
    niestemskiDistinctBosonicMode2007, chiScanningTunnelingSpectroscopy2012,
    shanEvidenceSpinResonance2012, zasadzinskiCorrelationTunnelingSpectra2001, ramiresEmulatingHeavyFermions2021}.
  Such features are usually interpreted as a signature of bosonic modes that
  mediate superconductivity and can thus provide key insight into the pairing
  mechanism\cite{carbottePropertiesBosonexchangeSuperconductors1990,
    songPairingInsightsIronbased2013}. If a superconductor exhibits strong
  electron-boson coupling, dip-hump signatures are expected to appear at
  energies $\Pi = \Delta + \Omega$, where $\Delta$ is the spectroscopic gap
  defined above and $\Omega$ is the bosonic-mode excitation
  energy\cite{scalapinoStrongCouplingSuperconductivity1966,
    carbottePropertiesBosonexchangeSuperconductors1990,songPairingInsightsIronbased2013}.
  We extract the energy of the mode $\Omega = \Pi - \Delta$ as a function of
  doping (\prettyref{fig: fig4}b) and find it to be correlated with $\Delta$. In
  the V-shaped region, $\Omega/(2\Delta)$ anticorrelates with the spectroscopic
  gap---in agreement with the trends seen in cuprates and iron-based
  compounds\cite{leeInterplayElectronLattice2006,
    niestemskiDistinctBosonicMode2007,zasadzinskiCorrelationTunnelingSpectra2001,shanEvidenceSpinResonance2012,yuUniversalRelationshipMagnetic2009}---and
  is bounded to be less than $1$ (\prettyref{fig: fig4}c). The upper bound of
  $\Omega/(2\Delta) \le 1$
  suggests\cite{andersonTheoryAsymmetricTunneling2006,eschrigEffectMagneticResonance2003,yuUniversalRelationshipMagnetic2009}
  that the pairing glue originates from a collective mode related to electronic
  degrees of freedom (see Refs.~\citenum{khalafChargedSkyrmionsTopological2021}
  and~\citenum{fischerUnconventionalSuperconductivityMagicAngle2021} for
  examples of such mechanisms), as electronic excitations with energy above
  $2\Delta$ become rapidly damped by the particle-hole continuum, unlike for
  phonon modes. We cannot, however, rule out low-energy ($<2\Delta$) phonons
  \cite{choiDichotomyElectronPhononCoupling2021} through this line of argument
  since higher-energy phonon dip-hump features may not be resolvable in our
  experiment. Even if not directly related to the pairing mechanism, dip-hump
  features anticorrelated with the gap may be valuable signatures of a proximate
  competing order, as discussed in relation to the
  cuprates\cite{reznikElectronPhononCoupling2006,
    letaconInelasticXrayScattering2014, gabovichChargeDensityWaves2014} or even
  in the context of twisted bilayer graphene
  \cite{caoNematicityCompetingOrders2021}. In the U-shaped region,
  $\Omega/(2\Delta)$ does not exhibit a clear anticorrelation with the
  spectroscopic gap, possibly due to subtleties with extracting the true
  superconducting order parameter in the BEC phase.

  Signatures of MATTG superconductivity presented in this work include: (i) coherence
  peaks that are suppressed with temperature and magnetic field, but persist
  well beyond the BCS limit; (ii) a pseudogap-like regime; (iii) dip-hump
  structures in the tunneling conductance; and (iv) tunneling conductance
  profiles that are not adequately fit with an $s$-wave order parameter, but
  instead are compatible with a gate-tuned transition from a gapped BEC to a
  gapless BCS phase with a common nodal order parameter. Parallel spectroscopic measurements on twisted bilayer graphene revealed similar phenomenology\cite{yazdaniSpectroscopicSignaturesNodal2021}---including nodal tunneling spectra, giant gap-to-$T_C$ ratios, and pseudogap physics with anomalous resilience to temperature and magnetic fields---suggesting a common origin of superconductivity in bilayers and trilayers. Properties (i-iii) are
  typically associated with non-phonon-mediated pairing, although phonon-driven
  mechanisms can exhibit some of these
  features\cite{lewandowskiPairingMagicangleTwisted2021,
    chouCorrelationinducedTripletPairing2021}. Regardless of pairing-mechanism
  details, together with property (iv), the observed signatures provide
  unambiguous spectroscopic evidence of the unconventional nature of MATTG
  superconductivity. Future theories addressing (i-iv) will likely be needed to
  pinpoint the exact mechanism of superconductivity in this system.


\noindent {\bf Acknowledgments:} We acknowledge discussions with Felix von
Oppen, Gil Refael, Yang Peng, and Ali Yazdani. {\bf Funding:} This work has been primarily
supported by Office of Naval Research (grant no. N142112635); National Science 
Foundation (grant no. DMR-2005129); and Army Research Office under Grant
Award W911NF17-1-0323. Nanofabrication efforts have been in part supported by
Department of Energy DOE-QIS program (DE-SC0019166). S.N-P. acknowledges 
support from the Sloan Foundation. J.A. and S.N.-P. also acknowledge support 
of the Institute for
Quantum Information and Matter, an NSF Physics Frontiers Center with support of
the Gordon and Betty Moore Foundation through Grant GBMF1250; C.L. acknowledges
support from the Gordon and Betty Moore Foundation’s EPiQS Initiative, Grant
GBMF8682. A.T. and J.A. are grateful for the support of the Walter Burke
Institute for Theoretical Physics at Caltech. H.K. and Y.C. acknowledge support
from the Kwanjeong fellowship.

\noindent {\bf Author Contribution:} H.K. and Y.C. fabricated samples with the
help of Y.Z. and R.P., and performed STM measurements. H.K., Y.C., and S.N.-P.
analyzed the data. C.L. and A.T. provided the theoretical analysis supervised by
J.A. S.N.-P. supervised the project. H.K., Y.C., C.L., A.T., J.A., and S.N.-P.
wrote the manuscript with input from other authors.

\noindent {\bf Data availability:} The data that support the findings of this
study are available from the corresponding authors on reasonable request.

\clearpage

\clearpage

\clearpage

\clearpage

\clearpage

\beginsupplement

\noindent {\bf Methods:}

\section{\bf Device fabrication} \label{methods: fabrication} Similarly as in
our previous STM measurements\cite{choiElectronicCorrelationsTwisted2019,
  choiCorrelationdrivenTopologicalPhases2021,
  choiInteractiondrivenBandFlattening2021} the device is fabricated using the
Polydimethylsiloxane (PDMS)-assisted stack-and-flip technique using $\sim$30nm
hBN and monolayer graphene. The flakes are exfoliated on SiO$_2$ and identified
optically. We use a poly(bisphenol A carbonate) (PC)/PDMS stamp to pick up hBN
at 90$\degree$C, and tear and twist graphene layers at 40$\degree$C. The PC film
with the stack is then peeled off and transferred onto another clean PDMS, with
MATTG side facing the PDMS. The PC film is dissolved in N-Methyl-2-pyrrolidone
(NMP), followed by cleaning with Isopropyl alcohol (IPA). We kept the final PDMS
in vacuum for several days. The stack on it is then transferred onto a chip with
a graphite back gate and gold electrodes. Finally, MATTG is connected to the
electrodes by another graphite flake.

\section{ \bf STM measurements} \label{methods: stm}

The STM measurements were performed in a Unisoku USM 1300J STM/AFM system using
a Platinum/Iridium (Pt/Ir) tip as in our previous works on
bilayers\cite{choiElectronicCorrelationsTwisted2019,
  choiCorrelationdrivenTopologicalPhases2021,
  choiInteractiondrivenBandFlattening2021}. All reported features are observed
with many (more than ten) different microtips. Unless specified otherwise, the
parameters for $dI/dV$ spectroscopy measurements were $V_{\rm Bias} = 100$~mV
and $I = 1$~nA, and the lock-in parameters were modulation voltage
$V_{\rm mod} = 0.2-1$~mV and frequency $f = 973$~Hz. The piezo scanner is
calibrated on a Pb(110) crystal by matching the lattice constant and verified by
measuring the distance between carbon atoms. The twist-angle uncertainty is
approximately $\pm0.01\degree$, and determined by measuring moir\'e wavelengths
from topography. Filling factor assignment has been performed by taking Landau
fan diagrams as discussed
previously\cite{choiCorrelationdrivenTopologicalPhases2021}.

\section{ \bf Stripe simulation} \label{methods: stripe}

As mentioned in the main text, the stripes are believed to arise out of the
restructuring of the moiré lattice. The flat-band scenario of interest arises
when the top and bottom monolayers are AA stacked---all carbon atoms vertically
aligned---while the middle layer is rotated by a twist angle $\sim1.5^\circ$.
While this situation seems understandably difficult to achieve during
fabrication, it was shown in
Ref.~\citenum{carrUltraheavyUltrarelativisticDirac2020} that the desired AA
stacking of the top and bottom is the energetically preferred configuration, and
we therefore expect the system to relax into this configuration across large
regions of the sample. This expectation is borne out by the observation of flat
bands as well as the presence of a single moiré lattice constant as seen in STM.

There are two primary features in \prettyref{fig: fig1}b that we wish to
reproduce. The first, and most prominent, is the stripe, which can be obtained
as follows. We let $\boldsymbol{a}_{1}=a(0,1)$ and
$\boldsymbol{a}_2=a(-\sqrt{3}/2,-1/2)$ denote the Bravais primitive vectors of
the graphene monolayer, where $a\approx \unit[0.246]{\text{nm}}$ is the graphene
lattice constant, and let $R(\phi)= e^{-i\phi \sigma^y}$ be a matrix that
rotates by angle $\phi$. The bottom and middle lattices are simulated by
plotting points at
$\Lambda_\mathrm{bot}=\big\{ R(-\theta/2)\big( n_1 \boldsymbol{a}_1+n_2 \boldsymbol{a}_2\big), n_{1,2}\in\mathbb{Z}\big\}$
and
$\Lambda_\mathrm{mid}=\big\{ R(\theta/2)\big( n_1 \boldsymbol{a}_1+n_2 \boldsymbol{a}_2\big), n_{1,2}\in\mathbb{Z}\big\}$.
The stripe is then entirely determined by strain in the top layer, where the
points plotted are instead
$\Lambda_\mathrm{top}=\big\{ R(-\theta/2)\big( n_1 \boldsymbol{a}_1+n_2 \boldsymbol{a}_2 + f_w(-\frac{1}{2}n_1+n_2)\boldsymbol{v}\big), n_{1,2}\in\mathbb{Z}\big\}$
where $f_w(x)=\frac{1}{\pi}{\arctan}(2x/w)+\frac{1}{2}$ and
$\boldsymbol{v}=m_1\boldsymbol{a}_1+m_2\boldsymbol{a}_2$, $m_{1,2}\in\mathbb{Z}$
is a Bravais lattice vector. The function $f_w(x)$ is essentially a smoothened
step function in that it interpolates between 0 and 1:
$\lim_{x\to-\infty}f_w(x)=0$ and $\lim_{x\to+\infty}f_w(x)=1$. The size of the
intermediate regime and hence the stripe width is determined by the parameter
$w>0$, with $\lim_{w\to0}f_w(x)=\Theta(x)$, the Heaviside function. In our
definition of $\Lambda_\mathrm{top}$, we chose to have $f_w$ as a function
$-\frac{1}{2}n_1+n_2$ since it results in a stripe along the $(-1/2,\sqrt{3}/2)$
direction and thus well represents the stripes shown in \prettyref{fig: fig1}b.
Putting these pieces together, one can see that in both regions where
$\left|-\frac{1}{2}n_1+n_2\right|$ is large, the lattice points of
$\Lambda_\mathrm{bot}$ and $\Lambda_\mathrm{top}$ should be directly above one
another. In the region $-w/2\lesssim -\frac{1}{2}n_1+n_2 \lesssim w/2$, the
registry of the top and bottom layers changes from AA to AB and then back to AA.

The procedure detailed above yields a stripe, but does not account for a second
feature of \prettyref{fig: fig1}b: the moiré lattices on either side of the
stripe are offset by about half of a moiré unit cell in the vertical
($\hat{\vec{y}}$) direction, which corresponds to a displacement of
$\vec{D}_\mathrm{shift}=\frac{\sqrt{3}}{2}L_M\hat{\vec{y}}$,
$L_M=a/\big(2\sin(\theta/2)\big)$. This offset at the moir\' e lattice scale is
a result of a shift of the top and bottom lattices relative to the middle
lattice occurring at the level of the microscopic scale of monolayer graphene.
In particular, displacing the top and bottom layers by
$\vec{v}_\mathrm{shift}\approx \theta\hat{\vec{z}}\times \vec{D}_\mathrm{shift}\approx-\frac{\sqrt{3}}{2}a\hat{\vec{x}}$
moves the moiré lattice by $\vec{D}_\mathrm{shift}$. Such a shift is readily
implemented numerically by replacing the lattices $\Lambda_\mathrm{bot}$ and
$\Lambda_\mathrm{top}$ with
$\Lambda_\mathrm{bot}'=\big\{ R(-\theta/2)\big( n_1 \boldsymbol{a}_1+n_2 \boldsymbol{a}_2+f_w(-\frac{1}{2}n_1+n_2)\boldsymbol{v}_\mathrm{shift}\big), n_{1,2}\in\mathbb{Z}\big\}$
and
$\Lambda_\mathrm{top}'=\big\{ R(-\theta/2)\big( n_1 \boldsymbol{a}_1+n_2 \boldsymbol{a}_2 + f_w(-\frac{1}{2}n_1+n_2)(\boldsymbol{v}+\vec{v}_\mathrm{shift})\big), n_{1,2}\in\mathbb{Z}\big\}$.
The middle layer is defined through $\Lambda_\mathrm{mid}$ as in the previous
paragraph. For ease of visualization, $\Lambda_\mathrm{top}'$ and
$\Lambda_\mathrm{bot}'$ are plotted in black while $\Lambda_\mathrm{mid}$ is
plotted in red.

We emphasize that the primary purpose of this calculation is to reproduce the
stripe in the simplest possible manner. A more complete study requires
understanding the energetics, which would not only be needed to predict that
width of the stripe (here, simply an input parameter), but which would also
result in lattice relaxation within a unit cell.

\section{ \bf Continuum model and Interaction-driven band structure renormalization} \label{methods: Hartree}

\subsection{Continnum model}

In this section, we summarize the continuum
model\cite{khalafMagicAngleHierarchy2019,carrUltraheavyUltrarelativisticDirac2020}
used to capture the low-energy theory of twisted trilayer graphene. In
particular, we consider the case where the top and bottom layers are directly
atop one another (AA stacked) and twisted by $-\theta/2$, while the middle layer
is twisted by $+\theta/2$. The electronic structure of MATTG is obtained by an
extension\cite{khalafMagicAngleHierarchy2019} of the continuum model developed
originally for twisted bilayer graphene
(TBG)\cite{bistritzerMoireBandsTwisted2011}. As in that case, there are two
independent sectors in the non-interacting limit distinguished by the valley $K$
and $K'$. Without loss of generality, we therefore focus on valley $K$ in this
section; the model relevant to valley $K'$ may be obtained in a straightforward
manner through time reversal. We let $\psi_t$, $\psi_m$, and $\psi_b$ denote the
spinors one obtains by expanding the dispersion of monolayer graphene about
valley $K$ for the top, middle and bottom layers, respectively. In terms of the
microscopic operators of the graphene monolayers, that means
$\psi_\ell(\vk)=f_\ell(\vk+\vec{K}_\ell)$, $\ell=t,m,b$. Importantly, as a
result of the twist, the $K$ points of the different layers are not the same.
The model is composed of a `diagonal' Dirac piece and an `off-diagonal'
tunneling piece accounting for the moir\'e interlayer coupling:
$H_\mathrm{cont}=H_D+H_\mathrm{tun}$. The Dirac term is broken up into three
components, one for each layer, with $H_D=H_t+H_m+H_b$ where
\begin{align}
    H_\ell &= \int_\vk \psi^\dag_\ell (\vk) h_{\theta_\ell}(\vk) \psi_\ell(\vk),
    &
    h_{\theta_\ell}(\vk) 
    &= 
    -v_0 e^{i\theta_\ell\sigma^z/2}
    \big(k_x \sigma^x + k_y\sigma^y \big)
    e^{-i\theta_\ell/2}.
\end{align}
Above, $\ell=t,m,b$ identifies the layers, $v_0\sim\unit[10^6]{\text{m/s}}$ is
the Fermi velocity of the Dirac cones of monolayer layer graphene, and
$\sigma^{x,y,z}$ are Pauli matrices acting on the A/B sublattice indices of the
spinors $\psi_\ell$. The angle $\theta_\ell$ indicates the angle by which each
layer is rotated, with $\theta_t=\theta_b=-\theta/2$ and $\theta_m=+\theta/2$.
The magic angle for this model occurs for $\theta\approx 1.5^\circ$, which is
related to the magic angle of TBG through a prefactor of $\sqrt{2}$:
$\theta=1.5^\circ \approx \sqrt{2}\times 1.05^\circ$. The origins of this
relation trace back to a similarity transformation that maps the MATTG continuum
model into one of a decoupled TBG-like band structure with an interlayer
coupling (to be discussed) multiplied by $\sqrt{2}$ and a graphene-like Dirac
cone. We refer to Ref.~\citenum{khalafMagicAngleHierarchy2019} for an in-depth
explanation of this relation.

We assume that tunneling only occurs between adjacent layers:
\begin{align}\label{app:Htun}
    H_\mathrm{tun}
    &=
    \sum_{j=1,2,3}\int_\vk \Big(
    \psi^\dag_t(\vk) + \psi^\dag_b(\vk) \Big)T_j \psi_m(\vk+\vq_j) + h.c.,
\end{align}
where the momenta shift and the tunneling matrices are given by
\begin{align}
    \vq_j
    &=
    \frac{4\pi}{3L_M}R\left(\frac{2\pi}{3}(j-1)\right)
    \begin{pmatrix} 0   \\  -1  \end{pmatrix},
    \notag\\
    T_j
    &=
    w_0+w_1\left(e^{-2\pi(j-1)i/3}\sigma^+ + e^{2\pi (j-1)i/3} \sigma^- \right)
\end{align}
with $R(\phi) = e^{-i\phi \sigma^y}$ is a $2\times2$ matrix acting on vector
indices, $L_M=a/[2\sin(\theta/2)]$, and $\sigma^\pm=(\sigma^x\pm i \sigma^y)/2$.
The tunneling strength is determined by the parameters $w_0$ and $w_1$; in this
paper we set $(w_0,w_1)=\unit[(55,105)]{\text{meV}}$. (Note that the conventions
used in this section are rotated by 90$^\circ$ relative to those of section~\ref{methods: stripe}.)

This model possesses a number of symmetries. We have already alluded to time
reversal, with which one may obtain the continuum model Hamiltonian
corresponding to the valley $K'=-K$. We therefore re-introduce a valley label,
writing $\psi_\ell\to \psi_{v,\ell}$ with $v=K,K'$. A number of spatial
symmetries are also present in this model, but for our purposes it is sufficient
to note that the model is invariant under rotations by $60^\circ$, under which
the spinors transform as
$\psi_\ell(\vk)\to \tau^x\sigma^xe^{2\pi i \tau^z\sigma^z/3}\psi_\ell\big(R(2\pi/6)\vk\big)$,
where $\tau^{x,y,z}$ are Pauli matrices acting on the (now suppressed) valley
indices.

To diagonalize the continuum model, we recall that the spinor operators
$\psi_\ell$ are not all defined about a common momentum point. Hence the
tunneling term in Eq.~\eqref{app:Htun} does not involve a momentum exchange of
$\vq_j$, but rather $K_{t}=K_b = K_m+\vq_j$ and $K_{t}'=K_b' = K_m-\vq_j$, which
differ by a moiré reciprocal lattice vector. We therefore define operators
$\Psi_{v,\ell}$ about a common momentum point for each valley through
$\Psi_{v,t/b}(\vk)=\psi_{v,t/b}(\vk)$ and
$\Psi_{K/K',m}(\vk)=\psi_{K/K',m}(\vk\pm\vq_1)$, where the $+$ ($-$) corresponds
to $K$ ($K'$) (the choice $\vq_1$ is arbitrary---$\vq_2$ and $\vq_3$ could be
equally chosen). Grouping the valley, layer, sublattice, and spin labels into a
single indice, $\Psi_\alpha$,
we can express $H_\mathrm{cont}$ in matrix form as
\begin{align}
    H_\mathrm{cont}
    &=
    \sum_{\vec{G},\vec{G}'}\int_{\vk\in\mathrm{mBZ}}
    \Psi^\dag_\alpha(\vk+\vec{G})h^{(\mathrm{cont})}_{\alpha,\vec{G};\alpha',\vec{G}'}(\vk)\Psi_{\alpha'}(\vk+\vec{G}');
\end{align}
$\vec{G}$ and $\vec{G}'$ are moiré reciprocal lattice vectors defined via
$\vec{G}=n_1\boldsymbol{\mathcal{G}}_1+n_2\boldsymbol{\mathcal{G}}_2$,
$n_{1,2}\in\mathbb{Z}$ where $\boldsymbol{\mathcal{G}}_1=\vq_2-\vq_1$ and
$\boldsymbol{\mathcal{G}}_2=\vq_3-\vq_1$. The integration over $\vk$ includes
only those momenta within the moiré Brillouin zone (mBZ).

\subsection{Interaction-driven band structure renormalization}

The presence of flat bands in MATTG necessitates the consideration of
interaction-driven band structure corrections. As demonstrated experimentally in
our previous work on twisted graphene
bilayers\cite{choiInteractiondrivenBandFlattening2021}, filling-dependent
interaction effects, specifically Hartree and Fock corrections, drastically
alter the electron dispersion. Here we incorporate only a Hartree
mechanism\cite{guineaElectrostaticEffectsBand2018,rademakerChargeSmootheningBand2019,
  goodwinHartreeTheoryCalculations2020, calderonInteractions8orbitalModel2020}
in the analysis. In TBG we found\cite{choiInteractiondrivenBandFlattening2021}
that the main role of the Fock correction, provided that one does not consider
the nature of the correlated states and the cascade, is to broaden the band
structure at the charge neutrality point ($\nu = 0$) and to counteract band
inversions at the zone center promoted by Hartree effects. For comparison with
the experiment presented in Fig.~\ref{fig: fig2}, where we focus only on
$\nu=\pm 4$, we can thus ignore Fock corrections as a first approximation.
Similar Hartree-driven band structure renormalizations were considered recently
in the literature\cite{fischerUnconventionalSuperconductivityMagicAngle2021,
  phongBandStructureSuperconductivity2021}, and our analysis together with the
experimental results are consistent with their conclusions.

We introduce Coulomb interaction into the system through
\begin{align}\label{eqn:CoulombInt}
    H_C&=
    \frac{1}{2}\int d^2\vr \,d^2\vr' \,
    \delta\rho(\vr)V(\vr-\vr')\delta\rho(\vr').
\end{align}
Here, $V(\vr)=e^2/(4\pi\epsilon |\vr|)$ is the Coulomb potential and
$\delta\rho(\vr)=\Psi^\dag(\vr)\Psi(\vr)-\rho_\mathrm{CN}(\vr)$, where
$\rho_\mathrm{CN}(\vr)=\langle \Psi^\dag(\vr)\Psi(\vr)\rangle_\mathrm{CN}$ is
the expectation value of the density at the charge neutrality point. The use of
$\delta\rho(\vr)$ instead of $\rho(\vr)$ in the interaction is motivated by the
expectation that the input parameters of the model $H_\mathrm{cont}$ already
include the effect of interactions at the charge neutrality point. Although
numerically expedient, this assumption is not strictly correct since the input
parameters in actuality refer to three independent graphene monolayers.
Nevertheless, for the purpose of making qualitative comparisons with
Fig.~\ref{fig: fig2}, we do not expect this distinction to be important. The
dielectric constant $\epsilon$ in the definition of $V(\vr)$ is used as a
fitting parameter; see section~\ref{methods:bandwidth_estimation} for details.

We study the effect of the interacting continuum model of MATTG through a
self-consistent Hartree mean-field calculation. Instead of solving the many-body
problem, we obtain the quadratic Hamiltonian that best approximates the full
model when only the symmetric contributions of $H_C$ are included, i.e., the
Fock term is neglected as explained above. Thus instead of
$H_\mathrm{cont}+H_C$, we study the Hamiltonian
\begin{align}\label{eqn:Hmf_hartree}
    H_\mathrm{MF}^{(\nu)}
    &=
    H_\mathrm{cont}
    + 
    H^{(\nu)}_\mathrm{H}
    -
    \frac{1}{2}\langle H_\mathrm{H}^{(\nu)} \rangle_\nu,
\end{align}
where $H_\mathrm{H}^{(\nu)}$ is the Hartree term at filling $\nu$,
\begin{align}
    H_\mathrm{H}^{(\nu)}
    &=
    \int_{\vk,\vk',\vq}V(\vq)
    \langle \Psi^\dag(\vk'+\vq)\Psi(\vk') \rangle_{\nu}
    \Psi^\dag(\vk)\Psi(\vk-\vq),
\end{align}
and the last term in Eq.~\eqref{eqn:Hmf_hartree} simply ensures there is no
double counting when one calculates the total energy. In the above equation,
$V(\vec{q})=2\pi e^2/(\epsilon |\vq|)$ is the Fourier transform of the Coulomb
interaction $V(\vr)$ in Eq.~\eqref{eqn:CoulombInt}, and the expectation value
$\langle\hat{\mathcal{O}}\rangle_\nu = \langle\hat{\mathcal{O}}\rangle_\mathrm{occ} - \langle\hat{\mathcal{O}}\rangle_\mathrm{CN}$
only includes states that are filled up to $\nu$ \emph{relative} to charge
neutrality, {as defined by diagonalizing the Hamiltonian
  $H_\mathrm{MF}^{(\nu)}$}. Typically, for a ``jellium''-like model, the
expectation value vanishes save for $\vq=0$, which is subsequently cancelled by
the background charge---allowing one to set $V(\vq=0)=0$ and completely ignore
the Hartree interaction. However, because the moiré pattern breaks continuous
translation symmetry, momentum is only conserved modulo a reciprocal lattice
vector. We therefore obtain
\begin{align}\label{eqn:Hartree1}
    H_\mathrm{H}^{(\nu)}
    &=
    \sum_{\vec{G}}'
    V(\vec{G})
    \int_{\vk'}
    \langle \Psi^\dag(\vk'+\vec{G})\Psi(\vk') \rangle_{\nu}
    \int_{\vk}
    \Psi^\dag(\vk)\Psi(\vk-\vec{G}),
\end{align}
where the prime above the summation over the moiré reciprocal lattice vectors
indicates that $\vec{G}=0$ is excluded. The self-consistent procedure begins by
assuming some initial value of $H_\mathrm{H}^{(\nu)}$ and diagonalizing the
corresponding mean-field Hamiltonian $H_\mathrm{MF}^{(\nu)}$ to obtain the Bloch
wavefunctions and energy eigenvalues. These quantities are then used re-compute
the expectation values that define $H_\mathrm{H}^{(\nu)}$ and thus
$H_\mathrm{MF}^{(\nu)}$. This process is repeated until one obtains the
quadratic Hamiltonian $H_\mathrm{MF}^{(\nu)}$ that yields the correlation
functions $\langle\cdot\rangle_\nu$ used in its definition.

It has further been
shown\cite{guineaElectrostaticEffectsBand2018,ceaElectronicBandStructure2019}
that the Hartree potential is dominated by the first `star' of moiré reciprocal
lattice vectors, which in our conventions corresponds to
$\vec{G}_n=R\big(2\pi(n-1)/6\big)\frac{4\pi}{\sqrt{3}L_M}(1,0)^T$ for
$n=1,\dots,6$, with $R(\phi)$ a rotation matrix. In this last approximation that
we employ, the $2\pi/6$ rotation symmetry of the continuum model greatly
simplifies the calculation of the Hartree term. Notably,
$V(\vec{G})\int_{\vk'}\langle\Psi^\dag(\vk'+\vec{G})\Psi(\vk')\rangle_\nu$ must
be the same for all $\vec{G}_n$, and, instead of Eq.~\eqref{eqn:Hartree1}, we
use
\begin{align}
\label{eq:sc_equations}
    H_\mathrm{H}^{(\nu)}
    &=
    V_\mathrm{H}^{(\nu)}
    \sum_{n=1}^6 \int_\vk \Psi^\dag(\vk)\Psi(\vk-\vec{G}_n),
    &
    V_\mathrm{H}^{(\nu)}
    &=
    \frac{1}{6}\sum_{n=1}^6V(\vec{G}_n)\int_{\vk'}\langle\Psi^\dag(\vk'+\vec{G})\Psi(\vk')\rangle_\nu\,.
\end{align}
The self-consistent procedure in this case is identical to that described in the
previous paragraph, but due to the reduced number of reciprocal lattice vectors
that are included in the summation the calculation is computationally easier.
Convergence is typically reached within $\sim 6$ iterations.

For clarity, all bands corresponding to different fillings plotted in
Fig.~\ref{fig: fig2}b have been shifted so that the Dirac points of the flat
bands always occur at the zero of the energy scale; it follows that the
(independent) graphene-like Dirac cone is then displaced in energy relative to
the fixed reference point of the flat bands for each filling. If this procedure
was not performed for clarity purposes, then the Hartree calculation would yield
band structures with a graphene-like Dirac cone fixed at one energy for all
fillings, but with shifted flat bands relative to it, as predicted in ab-initio
calculations\cite{fischerUnconventionalSuperconductivityMagicAngle2021}.

\subsection{Hartree correction and estimate of dielectric constant}
\label{methods:bandwidth_estimation}

As discussed in the previous section, due to Hartree corrections, the Dirac
cones shift downwards (upwards) in energy relative to the flat bands under
electron (hole) doping, as seen in \prettyref{fig: fig2}b-d. These relative
shifts are measured to be rather large ($\approx \unit[70]{\text{meV}}$ for
$\nu = +4$ and $\approx \unit[50]{\text{meV}}$ for $\nu= -4$), similar to the
bandwidth of the MATTG flat bands (approximately $50$~meV). These relative
shifts allow us to estimate an effective dielectric constant $\epsilon$ to be
used in Hartree band-structure-renormalization calculations. In particular, we
find that $\epsilon = 12-13$ quantitatively reproduces the observed Dirac point
shifts at $\nu = \pm 4$. Finally, we note that the relative shift between Dirac
cones and flat bands may also explain a certain discrepancy between our
measurements and the bandwidth estimates of the flat bands found in
transport\cite{parkTunableStronglyCoupled2021} that assumed fixed relative
position between Dirac point and flat points. This assumption, neglecting the
Hartree correction leads to an overestimate of a bandwidth by a factor of
$\sim 2$ (we measure flat band width to be approximately $50$~meV while Ref.
\citenum{parkTunableStronglyCoupled2021} found it to be around $100$~meV).

\section{ \bf Tunneling conductance normalization and fitting procedure} \label{methods: fitting}

In \prettyref{fig: fig3}b,c the tunneling conductance has been normalized by
dividing the spectra with a sixth-order polynomial fit that preserves the area
of the spectrum \cite{zasadzinskiTunnelingSpectroscopyBi2Sr2CaCu2O82003} (see
also \prettyref{exfig: fig9}). This procedure returns normalized dI/dV curves
that approach unity outside of the spectroscopic gap and removes in part the
large asymmetry between electrons and holes near $\nu = -2$ and above
$V_{\rm Bias} = 5$~meV. We emphasize that the regimes displaying U- and V-shaped
tunneling spectra are clearly visible both before and after this normalization
procedure. The dip-hump structure persists after this step as well (see black
arrow in \prettyref{exfig: fig9}).

The normalized dI/dV curves are fitted with the Dynes
formula\cite{dynesDirectMeasurementQuasiparticleLifetime1978},
\begin{equation}
\frac{d I}{d V}\propto \int_{-\infty}^{\infty} d\omega \int_0^{2\pi} d\theta ~\mathrm{Re}\left[\frac{\omega+i\Gamma}{\sqrt{(\omega+i\Gamma)^2-\Delta(\theta)^2}}\right] \left.\left(-\frac{d f}{d\omega}\right)\right\rvert_{\omega=\omega+eV}\,, %
\label{Dynes}
\end{equation}
where $f(\omega) = 1/(e^{\omega/k_B T}+1)$ ($k_B$ is a Boltzmann constant and
$T=400$~mK in our measurements); $\Delta(\theta)$ is the superconducting pairing
potential and; spectral broadening coming from disorder and finite lifetime of
Cooper pairs are incorporated by the parameter $\Gamma$. We consider isotropic
$s$-wave pairing, a pairing with a nodal order parameter, and a combination of
the two (see also section \ref{methods: shape} and \prettyref{exfig: fig7} for a
more detailed discussion and fits). For the nodal case we use
$\mathrm{\Delta}(\theta) = \Delta_{0}\cos(2\theta)$ (i.e., a $d$-wave profile),
though any $\mathrm{\Delta}(\theta) = \Delta_{0}\cos(N\theta)$ with integer
$N\neq 0$ gives the same spectrum. We therefore do not distinguish between
different nodal order parameters, e.g., $p$- versus $d$- versus $f$-wave. In the
plots, we also took into account the broadening due to finite lock-in modulation
excitation $V_{\rm mod} = 200\micro$~V.

\section{ \bf Possible Scenarios of U-shaped to V-shaped spectral evolution} \label{methods: shape}

In the main text, we introduced the experimental observation that the tunneling
conductance exhibits two qualitatively different tunneling profiles (U-
vs.~V-shaped) as a function of filling. We now discuss the details of two
possible scenarios for this outcome: $(i)$ a BCS-like superconductor with
filling-dependent order parameter symmetry and $(ii)$ a BEC-to-BCS transition
with a common nodal order parameter. As noted in the main text, we emphasize
that `BCS' in this context does \emph{not} imply any assumptions regarding the
pairing mechanism or coupling strength, but simply refers to a pairing scenario
wherein the chemical potential lies inside the band. Finally, we discuss the
Ginzburg-Landau coherence length in the BEC-BCS transition scenario and argue
that it is consistent with the results of
Ref.~\citenum{parkTunableStronglyCoupled2021}.

\subsection{BCS-like superconductor with filling-dependent order parameter symmetry}

The existence of U- and V-shaped tunneling spectra suggests that superconductivity  evolves with doping from a fully gapped to a gapless state. Here we address the possibility that these two regimes both arise from Cooper pairing a partially filled band with a Fermi surface, but with qualitatively different superconducting order parameters.   %
This scenario \emph{a priori} does not address the different behaviors of the Ginzburg-Landau coherence length $\xi_{\rm GL}$  seen in Ref.~\citenum{parkTunableStronglyCoupled2021}, e.g., the scaling of $\xi_{\rm GL}$ with the interparticle spacing 
(see section \ref{methods:GL_coherence_length}).  Nevertheless, whatever mechanism underlies the putative change in order parameter could potentially conspire to yield such dependence.  %

The V-shaped spectra can be adequately fit by postulating a nodal order parameter, as described in the main text and in section \ref{methods: fitting}.  In the present scenario, the U-shaped spectra are best fit by invoking multiple co-existing order parameters: either an $s$-wave gap together with a nodal order parameter or a combination of two nodal order parameters  (e.g., $d_{x^2-y^2}+i d_{xy}$) that together produce a gap in the tunneling conductance.  \prettyref{exfig: fig7}e displays the relevant fits. As noted in the main text, a similar change in pairing order with doping has been proposed in cuprates\cite{yehEvidenceDopingDependentPairing2001} (albeit with a less pronounced U-to-V evolution). Moreover, it has been argued that have argued that a $d_{x^2-y^2}+i d_{xy}$ spin-fluctuation-mediated pairing is energetically unfavourable compared to a real superposition of the two order parameters.\cite{fischerUnconventionalSuperconductivityMagicAngle2021}

\subsection{BEC-to-BCS transition}
\label{methods:bec_bcs_transition}

\subsubsection{Tunnelling current}

To describe the tunneling current expected in the BEC-BCS transition scenario and 
demonstrate qualitative consistency with experiment, we consider a 
phenomenological two-parabolic-band model. Specifically, we model the system 
near filling $\nu=-2$ with two bands of energy (in these two sections we set $\hbar = 1$)
\begin{equation}
      \xi_{\pm,\vec{k}}=\pm \left(\frac{k^2}{2m} + \Delta_\mathrm{CI}\right)-\mu,
\end{equation} 
separated by a $2\Delta_\mathrm{CI}$ correlated-insulator gap. Each band admits a two-fold `spin' degeneracy---which need not coincide exactly with physical spin, but could, e.g., represent some combination of spin and valley. In the absence of pairing, $\mu$ residing in the electron band $\xi_+$ (hole band $\xi_-$) corresponds to filling $\nu = -2+\delta n$ with $\delta n >0$ ($\delta n < 0$).  We focus primarily on the hole-doping case relevant for experiment.

For simplicity, we assume a `spin'-singlet, nodal $d$-wave pairing with a pair field $\Delta_\vec{k}$ that is the same in the electron and hole bands; inter-band pairing is neglected.  (We anticipate that triplet pairing would yield similar results, as would other nodal order parameters.)  The standard Bogoliubov–de Gennes formalism yields
\begin{align}
E_{\pm,\vec{k}}&=\sqrt{\xi_{\pm,\vec{k}}^2+\Delta_{\vec{k}}^2}\,,
&
u_{\pm,\vk}^2 &= 1+\frac{\xi_{\pm,\vec{k}}}{E_{\pm,\vec{k}}} \,, 
&
v_{\pm,\vk}^2 &= 1-\frac{\xi_{\pm,\vec{k}}}{E_{\pm,\vec{k}}}
\end{align}
with $u_{\pm,\vk}^2, v_{\pm,\vk}^2$ coherence factors describing overlap of the bare electron/hole wavefunctions with those of quasiparticles with dispersion $E_{\pm,\vec{k}}$.  The BEC phase corresponds to $|\mu|<\Delta_\mathrm{CI}$.  Here $\Delta_{\rm CI}$ renders the quasiparticles fully gapped despite the assumed nodal $d$-wave order parameter, and population of the electron and hole bands arises solely from pairing.   (At $\mu = 0$, the symmetry built into the electron and hole bands implies that the system remains undoped, corresponding to $\nu = -2$, even with $\Delta_{\vec k}\neq 0$.)  The regime $|\mu|>\Delta_{\rm CI}$ corresponds to a BCS phase wherein an electron- or hole-like Fermi surface undergoes Cooper pairing, yielding gapless quasiparticle excitations due to nodes in $\Delta_{\vec k}$.  Figure~\ref{fig: fig3}i,j schematically depicts the chemical potential associated with these two phases.  

The tunneling current follows from
\begin{equation}
\label{eq:tunneling_current}
   I(eV,\mu) \propto \sum_{s = \pm} \int d^2 \vk \, \left\{ u_{s,\vk}^2 \big[f(E_{s, \vec{k}}-eV) - f(E_{s,\vec{k}})\big]  - v_{s,\vk}^2 \big[1-f(-E_{s,\vec{k}}-eV)-f(E_{s,\vec{k}})\big]\right\},
\end{equation}
where $f(E) = 1/(e^{E/k_B T} +1)$ is the Fermi-Dirac distribution;
the differential tunneling conductance $d I/dV$ is obtained by numerically differentiating the current after the integral is evaluated. Below we will use this general formula to evaluate the tunneling conductance across the BEC-BCS transition.  As a primer, however, it is instructive to examine limiting cases.

Consider first the conductance deep in the BCS phase.  Here the current simplifies dramatically for relevant voltages.  First, focusing on the hole-doping case with $\mu \ll -\Delta_{\rm CI}$, we can neglect the electron band to an excellent approximation and focus solely on momenta near the Fermi surface for the hole band.  The remaining quasiparticle dispersion  $E_{-,\vec k}$ then has two `branches' with the same energy---corresponding to excitations above and below the hole-like Fermi surface (i.e., with $\xi_{-,\vec{k}} > 0$ and  $\xi_{-,\vec{k}} < 0$). 
That is, for each momentum $k^+>k_F$ ($k_F$ is the Fermi momentum), there exists a momentum $k^-<k_F$ such that 
$\xi_{-,\vec{k}^+}=-\xi_{-,\vec{k}^-}$, but $E_{-,\vec{k^+}} = E_{-,\vec{k^-}}$. The momentum-dependent part of the coherence factors therefore cancels, yielding a tunneling current
\begin{equation}
I(eV,\mu) \propto \int d^2 \vk \, \Big\{ \big[f(E_{-, \vec{k}}-eV) - f(E_{-,\vec{k}})\big]  - \big[1-f(-E_{-,\vec{k}}-eV)-f(E_{-,\vec{k}})\big]\Big\}
\end{equation}
that depends on the quasiparticle dispersion but not the coherence factors. Upon taking $d^2\vec{k}\approx k_F d k d\theta$, carrying out a variable change $\omega=\sqrt{\xi_{-,\vec{k}}^2+\Delta_{\vec{k}}}$, and assuming no $|\vec{k}|$ dependence in the pairing gap evaluated at the Fermi surface [$\Delta_{\vec{k}}\rightarrow \Delta(\theta)$], we arrive at the conventional BCS expression:
\begin{align}
I(eV,\mu) &\propto \int_0^{2\pi} d\theta \int_{\Delta(\theta)}^{\infty} d \omega \frac{\omega}{\sqrt{\omega^2-\Delta(\theta)^2}} \Big\{ \big[f(\omega-eV) - f(\omega)\big]  - \big[1-f(-\omega-eV)-f(\omega)\big]\Big\}\notag\\
&\propto \int_0^{2\pi} d\theta \int_{\Delta(\theta)}^{\infty} d \omega \frac{\omega}{\sqrt{\omega^2-\Delta(\theta)^2}} \left(-\frac{d f}{d \omega} eV\right)\notag\\
\implies \frac{d I}{d V} &\propto \int_0^{2\pi} d\theta \int_{\Delta(\theta)}^{\infty} d \omega \frac{\omega}{\sqrt{\omega^2-\Delta(\theta)^2}} \left(-\frac{d f}{d \omega}\right).
\end{align}
Implementing the Dynes substitution\cite{dynesDirectMeasurementQuasiparticleLifetime1978} $\omega\to\omega+i\Gamma$ then recovers the expression from Eq.~\eqref{Dynes}. The square-root factor in the denominator underlies coherence peaks associated with pairing-induced density-of-states rearrangement.  

By contrast, in the BEC phase ($|\mu| < \Delta_{\rm CI}$), or sufficiently close to the BEC-BCS transition, the simplifying procedure above breaks down.  Both electron and hole bands need to be retained; $\Delta_{\vec k}$ can not be simply evaluated at a Fermi surface, and hence dependence on the orientation \emph{and} magnitude of ${\bf k}$ become important; and since the minimum of the quasiparticle dispersion $E_{\pm,\vec{k}}$ occurs at or near $\vec{k} = 0$, the momentum-dependent part of the coherence factors no longer perfectly cancels.  
Together, these details manifest both through a ``softening'' of the coherence peaks in the tunneling conductance and the generation of a tunneling gap for \emph{any} pairing function $\Delta_{\vec{k}}$, $d$-wave or otherwise, in the BEC state; cf. Fig.~\ref{fig: fig3}k,l. 

Returning to the general current formula in Eq.~\eqref{eq:tunneling_current}, in simulations of Fig.~\ref{fig: fig3}k,l and supplemental simulations below, we employ a $d$-wave pairing potential with 
\begin{equation} 
\Delta_{\vk} = \Delta_0 h(k) \cos(2\theta).
\end{equation}
Here $k$ and $\theta$ are the magnitude and polar angle of $\vec{k}$, while $\Delta_0$ sets the pairing energy scale.  We take the $k$-dependent prefactor to be $h({k}) = \mathrm{tanh}({k}^2 \ell^2)$, where $\ell$ is roughly the real-space distance over which the $d$-wave pairing potential acts.  This choice results in $\Delta_{\vec k}$ vanishing at $k = 0$ as required for $d$-wave pairing, and regularizes the unphysical divergence that would appear with a simple $h(k) \propto k^2$ profile in a manner that preserves locality in real-space.  Let $\eta \equiv 2m\Delta_0 \ell^2$  be a dimensionless quantity involving $\ell$.  In the regime of the BCS phase with $k_F \ell  \gg 1$, near the Fermi surface we have $\Delta_{\vec k} \approx \Delta_0 \cos(2\theta)$; hence the value of $\eta$ is largely irrelevant provided $k_F^2/2m$ remains sufficiently large compared to $\Delta_0$.  In both the BCS regime with $k_F \ell \lesssim 1$ and throughout the BEC phase, the choice of $\eta$ is more significant.  Here, for the physically important `small' momenta, the pairing behaves like $\Delta_{\vec k} \approx \Delta_0 k^2 \ell^2 \cos(2\theta)$ and should be compared to the $k^2/2m$ kinetic energy scale.  With $\eta \lesssim 1$, pairing effects are suppressed since the latter scale dominates over the former.  By contrast, with $\eta \gtrsim 1$ the pairing scale dominates and correspondingly yields more dramatic signatures in density of states and tunneling conductance.  In particular, the coherence peaks appear most prominently in the BEC phase at $\eta \gg 1$. 

The tunneling conductance in the BEC and BCS phases can be studied as a 
function of chemical potential or as a function of filling. In our formalism, 
treating $\mu$ as the tuning parameter is more convenient since all $\mu$ 
dependence is contained in the quasiparticle dispersion $E_{\pm,\vec{k}}$ 
and the relation between filling and $\mu$ evolves nontrivially between 
the BEC and BCS phases.  In  experiment, however, the gate-controlled 
filling $\nu$ is the natural tuning parameter.  Additionally, the pairing 
strength and $\nu = -2$ gap, modeled here by $\Delta_0$ and $\Delta_{\rm CI}$, 
certainly depend on $\nu$---which further complicates the relation 
between filling and $\mu$. We defer a careful examination of this 
relation to future work.  Instead, here we will simply explore the 
tunneling conductance as a function of $\mu$, with $\mu$-dependent 
$\Delta_0$ and $\Delta_{\rm CI}$ input parameters extracted 
(crudely) from the experiment as follows.  

First, for each filling we fix $\Delta_0$ to the measured location of coherence peaks in Fig.~\ref{fig: fig3}h (and linearly extrapolate to continue to more negative $\mu$ values). In the V-shaped regime this assignment is expected to be quantitatively reliable, given our interpretation of that regime as a BCS phase (which would indeed have coherence peaks set by $\Delta_0$). However, the U-shaped regime, interpreted as a BEC phase, would have coherence peaks at an energy determined by multiple parameters including $\mu,\Delta_{\rm CI}$, and $\Delta_0$; thus here the assignment becomes an approximation that we invoke for simplicity.  We then obtain a $\Delta_0$ vs.~$\mu$ profile by naively replacing filling (or gate voltage) with $\mu$; i.e., we ignore the nontrivial relation linking these quantities. To determine $\Delta_{\rm CI}$ vs.~$\mu$, we first fix the value at $\mu = 0$ to be $\Delta_{\rm CI,0} = 2.7$~meV, corresponding to the $\nu = -2$ spectral gap seen in \prettyref{exfig: fig4}. 
We also fix the chemical potential $\mu_*$ corresponding to the BEC-BCS transition, which in our model occurs when $-\mu_* = \Delta_{\rm CI}(\mu_*)$.  We specifically set $\mu_* = -0.8$~meV so that the transition coincides roughly with the experimentally observed U-to-V change in \prettyref{fig: fig3} (after replacing density as $\mu$ as described above). 
We phenomenologically model the remaining $\mu$ dependence of $\Delta_{\rm CI}$ as
\begin{equation}
    \Delta_{\rm CI}(\mu) = \begin{cases} 
      \Delta_{\rm CI, 0} \frac{\gamma_{\rm CI}^2}{\mu^2+\gamma_{\rm CI}^2} & \mu \geq \mu_{*} \\
      \alpha_2 \mu^2+\alpha_1 \mu+\alpha_0 & \mu_{*} \geq  \mu 
   \end{cases}
\end{equation}
with $\alpha_2 = \Delta_{\rm CI}(\mu^+)/(\mu_{*}-\mu_{**})^2$, $\alpha_1 = - 2 \Delta_{\rm CI}(\mu^+) \mu_{**}/(\mu_{*}-\mu_{**})^2$, $\alpha_0 = \Delta_{\rm CI}(\mu^+) \mu_{**}^2/(\mu_{*}-\mu_{**})^2$ and $\mu_{**}=-1.1$ meV.  We further choose small enough $\gamma_{\rm CI} = 0.1$~meV to ensure coherence peak separation comparable with the experiment.  The parametrization above causes $\Delta_{\rm CI}$ to decrease upon hole doping and eventually vanish at a chemical potential $\mu_{**}$ (we fix $\Delta_{\rm CI}$ to zero beyond this point rather than allowing it to become negative).  This collapse of $\Delta_{\rm CI}$ is invoked to emulate experiment; $\mu$-independent $\Delta_{\rm CI}$ would produce additional structure in the tunneling conductance that is not resolved in measurements. 
Extended Data Fig.~\ref{fig:theory_figure_methods}a illustrates the resulting $\mu$ dependence of $\Delta_0$ and $\Delta_{\rm CI}$. 

Given these parameters, we evaluate the bias voltage and $\mu$ dependence of the 
tunneling conductance assuming $1/2 m\ell^2 =6.25~\mu$eV, which yields values 
of $\eta$ as large as ${\sim} 250$.  Extended Data Fig.~\ref{fig:theory_figure_methods}b,c 
presents tunneling conductance color maps and linecuts; data from 
\prettyref{fig: fig3}k,l were generated from the same parameter set.  While we caution against direct comparison of Fig.~\ref{fig: fig3}a and Extended Data Fig.~\ref{fig:theory_figure_methods}b given the crude model and parameter extraction used for the latter, our simulations do robustly capture the observed U- to V-shaped evolution.  Improved modeling of experiment could be pursued in several ways, e.g., by self-consistently relating $\mu$ and filling, and by employing more sophisticated band-structure modeling that accounts for density of states features at $\nu = -2$.  The latter in particular may be required to obtain more refined agreement with experimental details such as the relative coherence peak heights in the U- and V-shaped regimes.

\subsubsection{Connection to coherence length measurements}
\label{methods:GL_coherence_length}

Finally, we discuss the behaviour of the Ginzburg-Landau coherence length $\xi_\mathrm{GL}$ in the proposed BEC-BCS transition scenario.
The primary intent of this analysis is to emphasize that this scenario is consistent with the transport-based observations of Ref.~\citenum{parkTunableStronglyCoupled2021}, which found that $\xi_\mathrm{GL}$ admits two distinct regimes.
First, in the part of the superconducting dome with $\nu \lesssim -2.5$---roughly our V-shaped region---$\xi_{\rm GL}$ significantly exceeds the inter-particle spacing $d = 1/\sqrt{|\delta n|}$ (where $\delta n$ is measured relative to $\nu = -2$).
In this regime, the coherence length can be well captured by a standard form $\xi_{\rm GL} = c v_F/\Delta$ expected from dimensional analysis in a BCS phase, where $v_F$ is the Fermi velocity, $\Delta$ is the characteristic pairing energy, and $c$ is a (presumably order-one) constant.  Using $v_F \sim\unit[10^5]{\text{m/s}}$ (comparable to the flat-band velocity extracted from previous MATBG measurements\cite{choiCorrelationdrivenTopologicalPhases2021}), our measured spectroscopic gaps $\Delta$ (see above in section \ref{methods: fitting}), and $c \approx 2/3$ indeed yields coherence lengths that quantitatively agree with Ref.~\citenum{parkTunableStronglyCoupled2021} over this filling range.  For example, our measured $\Delta$ at $\nu = -2.5$ yields $\xi_{\rm GL} \approx \unit[30]{\text{nm}}$. 
This agreement supports the emergence of a `BCS' regime---albeit of a strongly coupled nature as confirmed by the anomalously large $2\Delta/(k_BT_C)$ ratio reported in the main text.

By contrast, in the complementary part of the superconducting dome with $\nu \gtrsim -2.5$---coinciding roughly with our U-shaped region---Ref.~\citenum{parkTunableStronglyCoupled2021} measured $\xi_\mathrm{GL}$ values that closely track the relative inter-particle spacing $d$ and become as small as $\sim \unit[12]{\text{nm}}$.  
The deviation from the form $\xi_\mathrm{GL}\propto v_F/\Delta$ can be accounted for by the presence of an additional energy scale, the gap for dissociating the Cooper-pair molecules, as well as the fact that $v_F$ has no meaningful definition in the absence of a Fermi surface.
Instead, the scaling relation $\xi_{\rm GL} \propto d$ is predicted for a BEC regime in related contexts\cite{pistolesiEvolutionBCSSuperconductivity1996, randeriaCrossoverBardeenCooperSchriefferBoseEinstein2014, stintzingGinzburgLandauTheorySuperconductors1997}, and we briefly sketch how the pertinent scaling may be obtained using the results of Ref.~\citenum{pistolesiEvolutionBCSSuperconductivity1996}.
We emphasize, however, that direct use of this reference requires a number of simplifying assumptions that limit the scope and applicability of the analysis.
Although the arguments outlined in the previous subsection hinge on the assumption of a nodal order parameter, we specialize here to nodeless $s$-wave pairing.
Nevertheless, because the BEC phase is gapped regardless of the function form of the gap, we do not expect this distinction to alter the functional relationship of $\xi_\mathrm{GL}$ vis-à-vis the interparticle spacing $d=1/\sqrt{|\delta n|}$.
We also restrict our attention to the hole band, $\xi_{-,\vk}$, which can be viewed as taking the $\Delta_\mathrm{CI}\to\infty$ limit in the model presented in the previous subsection. 
For convenience, we drop the subscript `$-$' as well as the reference to $\Delta_\mathrm{CI}$, simply expressing the dispersion as 
$\xi_\vk \equiv \xi_k= -k^2/(2m)-\mu$, where $k$ is the magnitude of  the vector $\vk$. 
It follows that $\mu>0$ corresponds to the BEC regime, while $\mu<0$ is the BCS regime (which we do not consider here). 
As in the previous subsection, details of the symmetry breaking leading to the $\nu=-2$ insulator are neglected, 
and a generic two-fold `spin' symmetry with quantum numbers labelled by $a=1,2$ is assumed to remain. 
A filling $\delta n$ of the hole bands corresponds to a filling $\nu=-2+\delta n$ of the TTG system with $\delta n<0$.

We start with a Hamiltonian
\begin{align}
    H
    &=
    \sum_{\vk,a}c^\dag_{a}(\vk)\xi_\vk c_{a}(\vk)
    +
    \sum_{\vk,\vk',\vq}Uc_1^\dag(\vk+\vq/2)c_2^\dag(-\vk+\vq/2)c_2(-\vk'+\vq/2)c_1(\vk'+\vq/2),
\end{align}
where $U$ characterizes the interaction strength and $c_{a = 1,2}(\vk)$ are electron annihilation operators.
The superconducting gap $\Delta$ that develops should be obtained from $H$ via a self-consistent equation, but for simplicity, we instead consider $\Delta$ as a constant, implying a superconducting spectrum given by $E_k=\sqrt{\xi_k^2+\Delta^2}$.
The macroscopically based coherence length $\xi_\mathrm{GL}$ is proportional to the microscopically derived $\xi_\mathrm{phase}$, which is identified with the inverse mass of the canonical boson $\phi(\vr)\sim c_1(\vr)c_2(\vr)$ in the effective action determined in Ref.~\citenum{pistolesiEvolutionBCSSuperconductivity1996}.
They find that $\xi_\mathrm{phase}=\sqrt{b/a}$ where
\begin{align}\label{eqn:xiPhase-ab-def}
    a&=\frac{\Delta^2}{4\pi}
    \int_0^\infty dk\,k\,\frac{1}{E_k^3}\cCom
    &
    b&=
    \frac{1}{32\pi m}
    \int_0^\infty dk \, k\,\frac{\xi_k^2}{E_k^5}
    \left[ 
    -\frac{\xi_k^2-2\Delta^2}{\xi_k}
    +
    \frac{5\Delta^2}{2m}\frac{k^2}{E_k^2}\right].
\end{align}
The model is analytically tractable, returning
\begin{align}
    \xi_\mathrm{phase}
    &=
    \sqrt{
    \frac{1}{12m}
    \frac{1}{x-\mu} 
    \left(
    \frac{\mu^2}{x^2}
    +
    \frac{x}{x+\mu}
    \right)},
    &
    x&=\sqrt{\mu^2+\Delta^2}.
\end{align}
This expression is explicitly a function of $\mu$ and not of the density $\delta n$ of the bands.
We relate the two via
\begin{align}
    \delta n&=
    -\frac{1}{2\pi}
    \int_0^\infty dk\,k\left(1+\frac{\xi_k}{E_k}\right),
\end{align}
which can be solved and inverted to obtain $\mu$ as a function of $\delta n$:
\begin{align}
    \mu
    &=
    \frac{(2\pi \delta n/m)^2-\Delta^2}{4\pi \delta n/m}
    \cdot
\end{align}
Deep in the BEC regime with $\delta n\to0^-$, we find
\begin{align}\label{eqn:xi_phase_BEC}
    \xi_\mathrm{phase}
    &\xrightarrow{ \, \delta n \to 0^-\, }
    \frac{1}{4\sqrt{-\pi \delta n}}
    \propto 
    d,
\end{align}
consistent with the observations of Ref.~\citenum{parkTunableStronglyCoupled2021}.
Hence, when comparing with experiment, $\xi_\mathrm{phase}$ has the 
same functional dependence on $d=1/\sqrt{|\delta n|}$ in the BEC regime.
Again, we emphasize that while the coefficient may differ, we do not expect 
the presence of nodes in the superconducting order parameter to alter 
our conclusions in this limit. 

We now turn to the intermediate regime between the BCS and BEC limits.
Based on transport measurements, Ref.~\citenum{parkTunableStronglyCoupled2021} proposed that MATTG can be tuned close to the BEC-BCS crossover (see also Ref.~\citenum{haoElectricFieldTunable2021}). We advocate for a complementary scenario, wherein the presence of gapless modes in the BCS regime implies that the system undergoes a 
BEC to BCS \emph{phase transition}.
This distinction was explicitly emphasized in Refs.~\citenum{botelhoLifshitzTransitionWave2005}  
in the context of the cuprates,
and the corresponding transition was also explored in Refs.~\citenum{stintzingGinzburgLandauTheorySuperconductors1997} and~\citenum{caoBCSBECQuantumPhase2013}.
The prospect of a gate-tuned transition within the superconducting dome 
is especially encouraging since it may be consistent with the apparent 
discontinuity in the coherence length measured in 
Ref.~\citenum{parkTunableStronglyCoupled2021}.
We leave the determination of the coherence length across the transition for future 
work.

\clearpage

\begin{figure}[p]
\begin{center}
    \includegraphics[width=15cm]{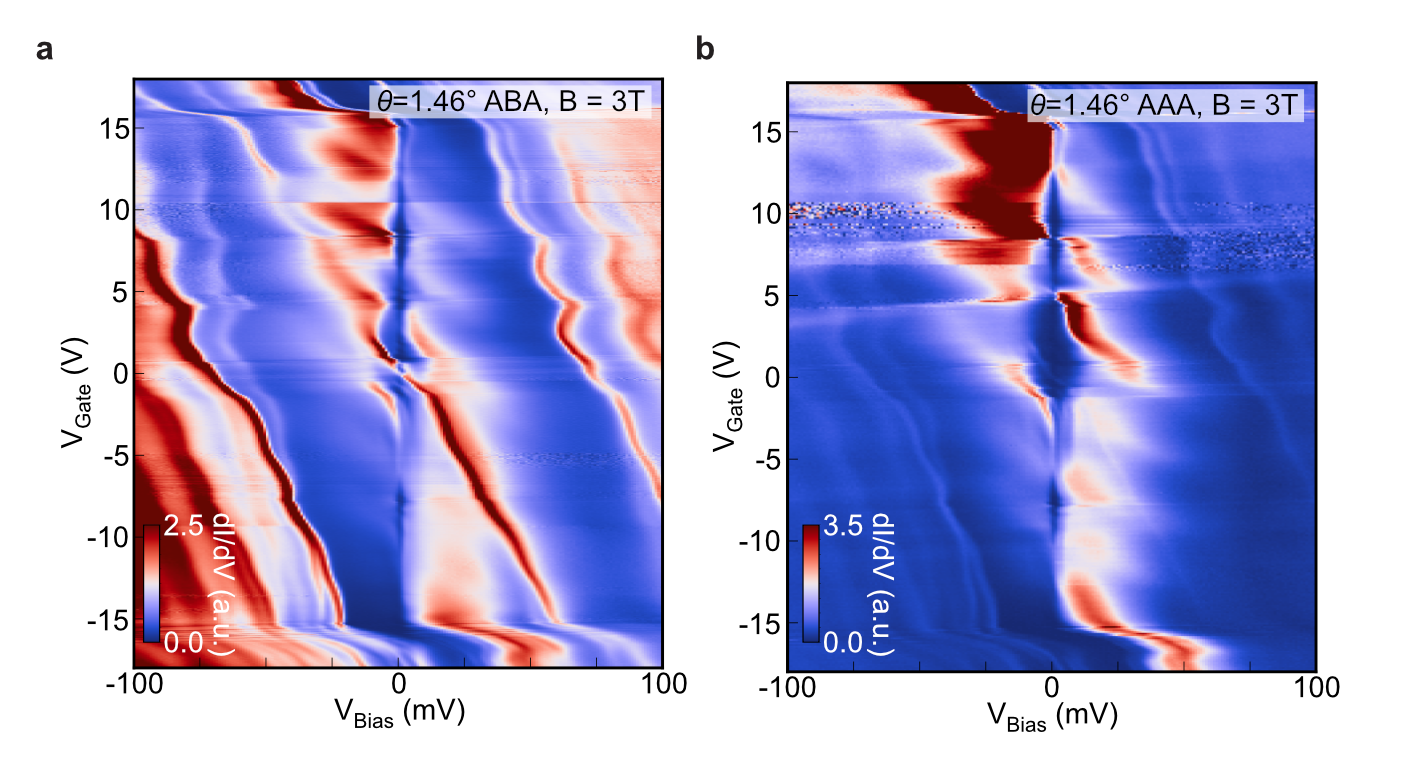}
\end{center}
\caption{{\bf Comparison between spectra on ABA and AAA sites at finite fields.} 
{\bf a-b}, Point spectroscopy as a function of $V_{\rm Gate}$ on 
ABA stacked ({\bf a}, the same as panel 
\prettyref{fig: fig2}d) and on AAA stacked ({\bf b}) region ($B = 3~\text{T}$, 
$\theta = 1.46\degree$). In comparison, flat bands appear to be more prominent on 
the AAA site ({\bf b}), while LLs from Dirac-like 
dispersion and dispersive bands appear more pronounced at ABA site. 
This is a direct consequence of LDOS from the flat bands being localized on the AAA 
sites. The LDOS from Dirac-like bands is spatially uniformly distributed. 
}
\label{exfig: fig2}
\end{figure}
\clearpage

\begin{figure}[p]
\begin{center}
    \includegraphics[width=15cm]{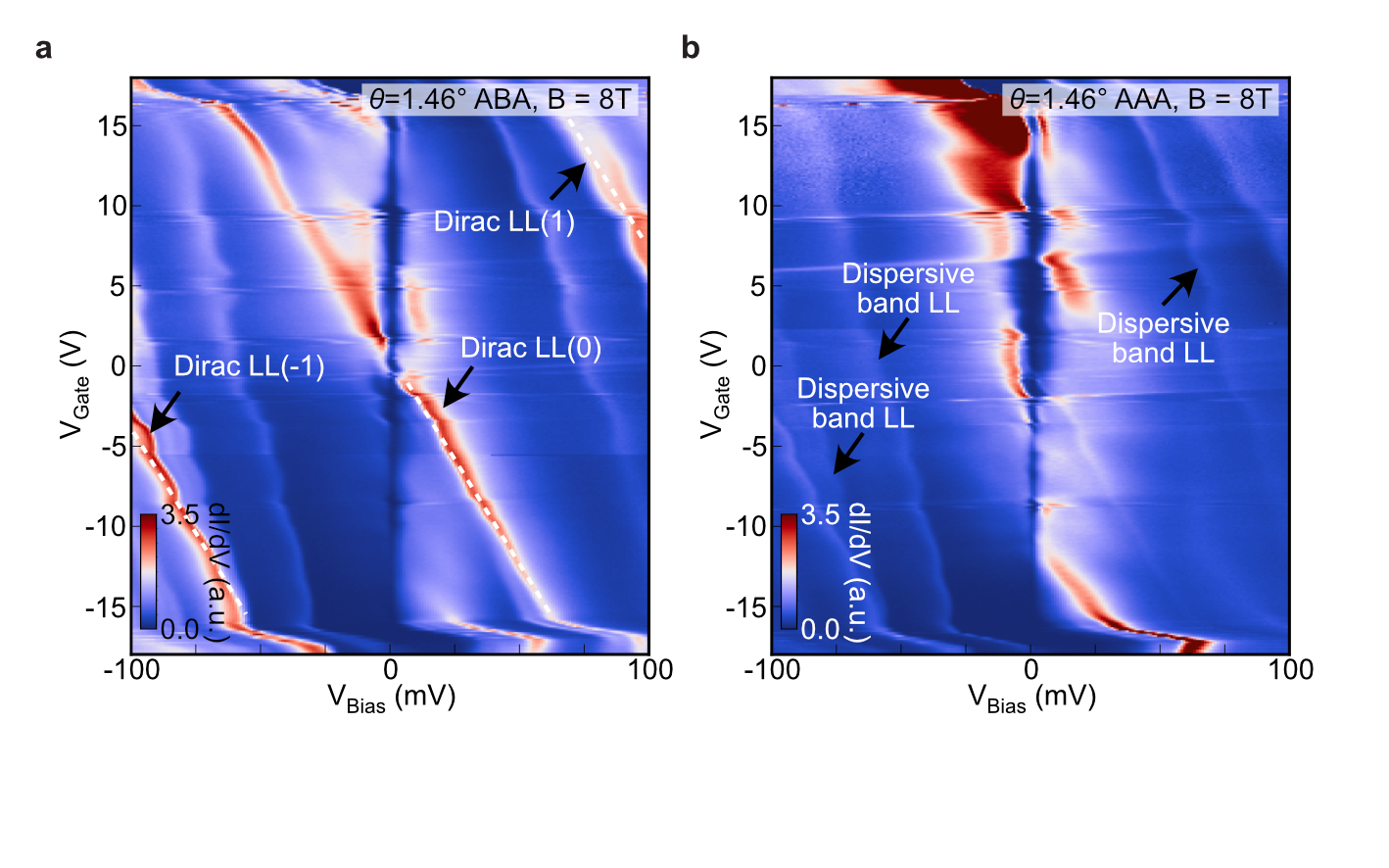}
\end{center}
\caption{{\bf Distinguishing dispersive band LLs and Dirac band LLs} {\bf a-b}, 
Point spectroscopy as a function of $V_{\rm Gate}$ on ABA stacked 
({\bf a}) and AAA stacked ({\bf b}) region ($B = 8~\text{T}$, 
$\theta = 1.46\degree$). Zeroth LL from Dirac dispersion is clearly
distinguished from other LLs as it crosses the flat band. Other LLs from Dirac 
dispersion is distinguished from the dispersive band from being parallel to 
the zeroth LL as a function of doping. Additional LL is observed at this high 
magnetic field at $V_{\rm Gate}>12~\text{V}$ which is more pronounced at AAA 
stacked region and can be attributed to second Dirac cone due to finite 
displacement field present at these $V_{\rm Gate}$.}
\label{exfig: fig3}
\end{figure}
\clearpage

\begin{figure}[p]
\begin{center}
    \includegraphics[width=14cm]{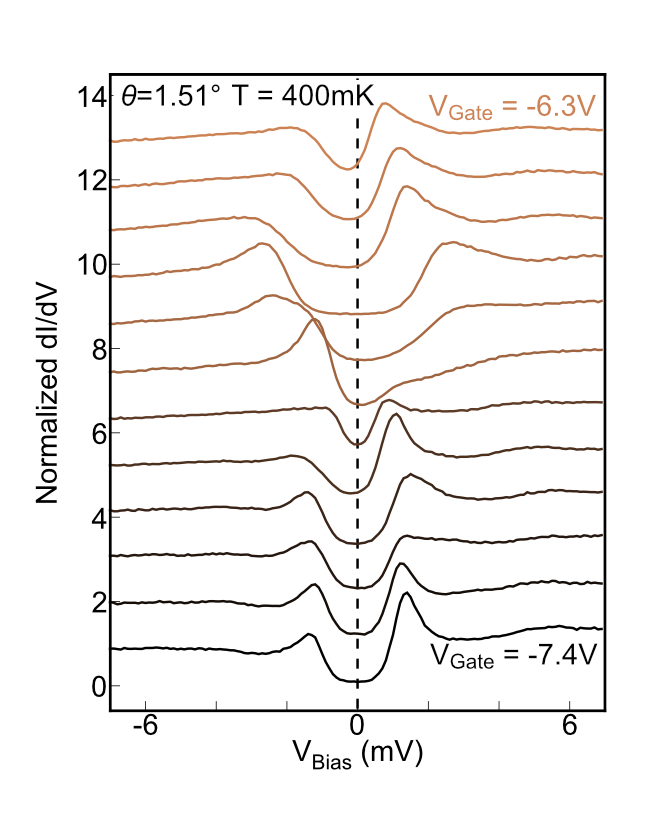}
\end{center}
\caption{{\bf Spectroscopy near $\nu = -2$.} Linecuts taken from 
\prettyref{fig: fig3}a for $V_{\rm Gate}$ ranging from 
$-6.3$~V to $-7.4$~V in $100$~mV steps. Starting from top, the 
observed gap is highly asymmetric and gradually evolves to the 
more symmetric spectrum on the bottom. Vertical dashed 
line shows the position of $V_{\rm Bias} = 0~\text{mV}$. We interpret 
that asymmetric gap (brown lines) corresponds to correlated insulator 
regime, while the symmetric gap (black lines) 
indicates superconducting regime.  
}
\label{exfig: fig4}
\end{figure}
\clearpage

\begin{figure}[p]
\begin{center}
    \includegraphics[width=14cm]{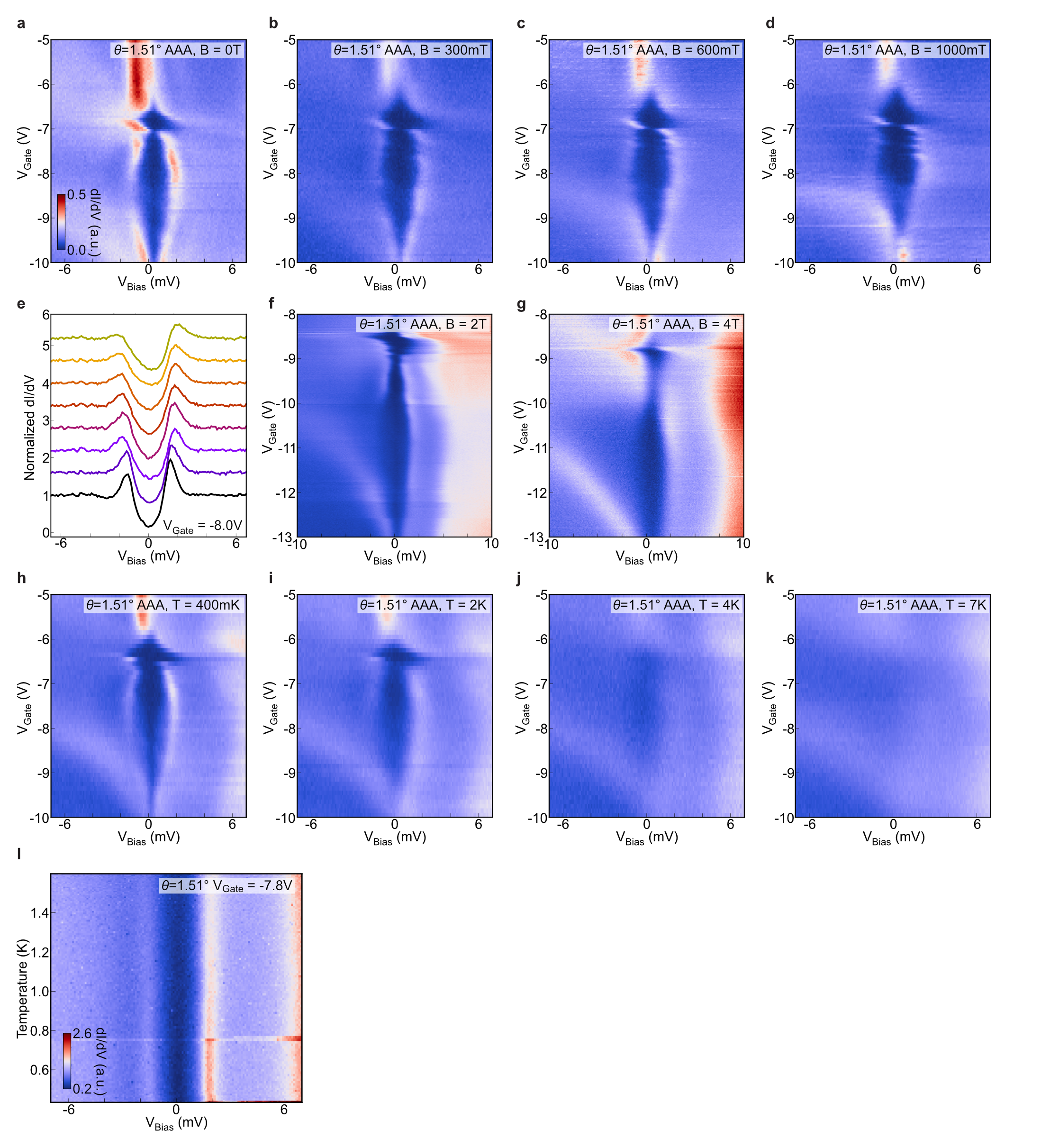}
\end{center}
\caption{{\bf Additional data sets showing magnetic field and temperature 
dependence of spectroscopic gap in the $\mathbf{-3<\nu<-2}$ range.} 
 {\bf a-d}, 
Point spectroscopy as a function of $V_{\rm Gate}$ at twist angle 
of $\mathrm{\theta = 1.51\degree}$ at magnetic field $\mathrm{B = 0}$~T 
({\bf a}), $\mathrm{B = 300}$~mT ({\bf b}), $\mathrm{B = 600}$~mT ({\bf c}), 
$\mathrm{B = 1}$~T ({\bf d}). {\bf e}, Line traces showing magnetic field 
dependence for $\mathrm{V_{Gate} = -7.8}$~V (U-shaped regime). 
Color coding corresponds to magnetic field $\mathrm{B = 0}$, $0.1$, 
$0.2$, $0.3$, $0.4$, $0.4$, $0.8$, $1$~T. Plots are offset for clarity. 
{\bf f, g}, Gate spectroscopy measured at $B = 2$~T ({\bf f}) and $B = 4$~T 
({\bf g}), for $\theta = 1.54\degree$ featuring gapped spectrum 
persisting $B \gtrsim 4$~T 
(data taken at different point compared to {\bf a-e}). {\bf h-k}, Gate
spectroscopy taken at different temperatures $\mathrm{T = 400}$~mK ({\bf h}), 
$\mathrm{T = 2}$~K ({\bf i}), $\mathrm{T = 4}$~K ({\bf j}), 
$\mathrm{T = 7}$~K ({\bf k}). {\bf i}, Point spectroscopy measured as a 
function of $V_{\rm Bias}$ and temperature at the same point as 
({\bf h-k}) for ${V_{\rm Gate} = -7.8}$~V. }
\label{exfig: fig5}
\end{figure}
\clearpage

\begin{figure}[p]
\begin{center}
    \includegraphics[width=14cm]{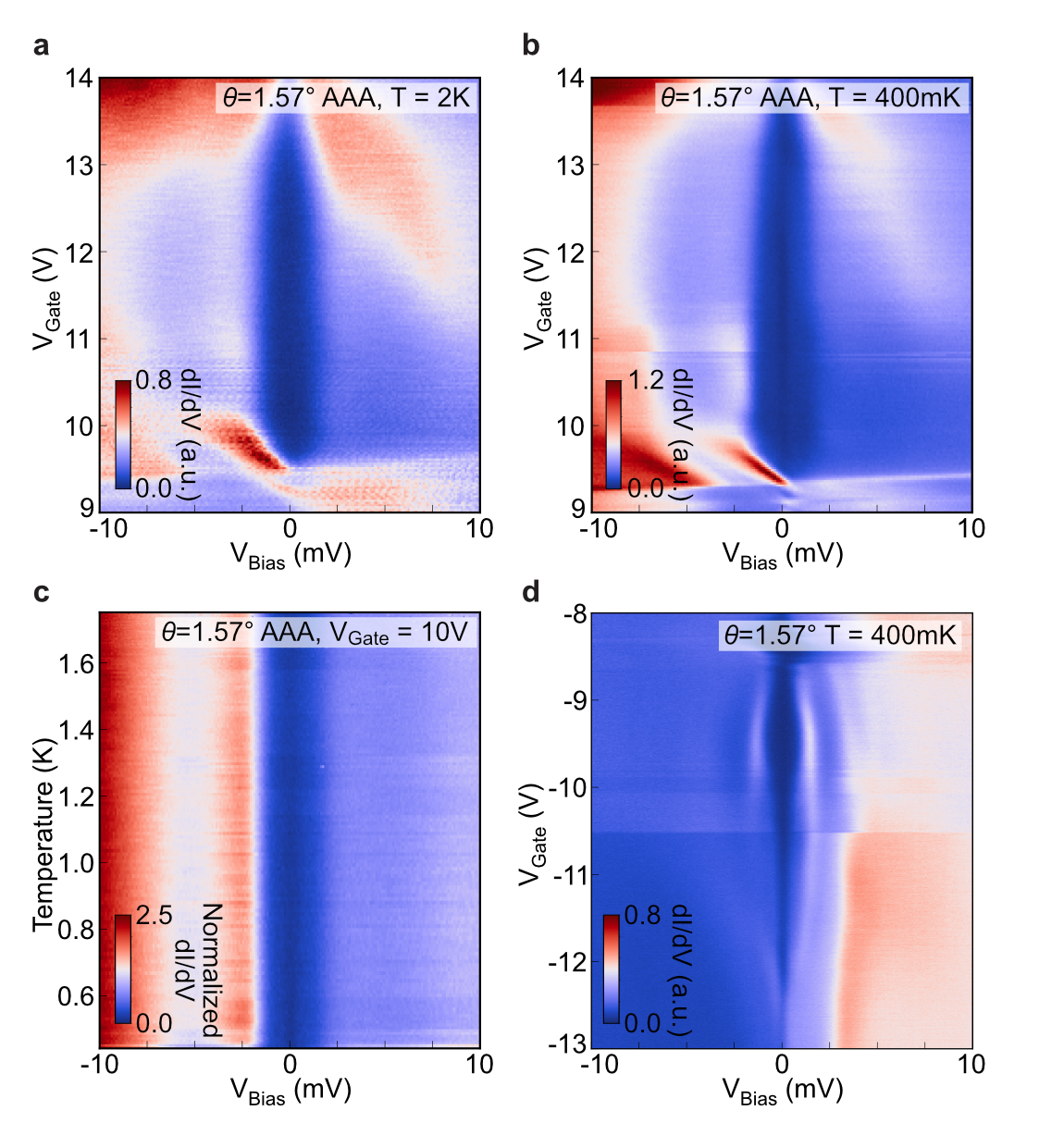}
\end{center}
\caption{{\bf Spectroscopic gap in the $+2<\nu<+3$ range.} 
{\bf a}, Tunneling conductance spectroscopy at twist angle of
$\mathrm{\theta = 1.57\degree}$ on AAA stacked region 
at $\mathrm{T = 2}$~K showing well-developed gapped region on the 
electron-side. {\bf b}, Spectroscopy 
measured at the same region at $\mathrm{T = 400}$~mK. {\bf c}, Spectroscopy
as a function of temperature at the same point as ({\bf a, b}) for 
$\mathrm{V_{Gate} = 10V}$. {\bf d}, Spectroscopy focusing on hole doping taken
with the same micro-tip. While the spectrum for hole doping ({\bf d}) shows clear 
coherence peaks and dip-hump structures these features are absent for the 
gap on the electron-side. We speculate that for electron doping, the coherence peaks
are suppressed even at our base temperature ($\mathrm{T = 400}$~mK). The observed
gap in this case is likely originating from pseudogap phase.}
\label{exfig: fig6}
\end{figure}
\clearpage

\begin{figure}[p]
\begin{center}
    \includegraphics[width=16cm]{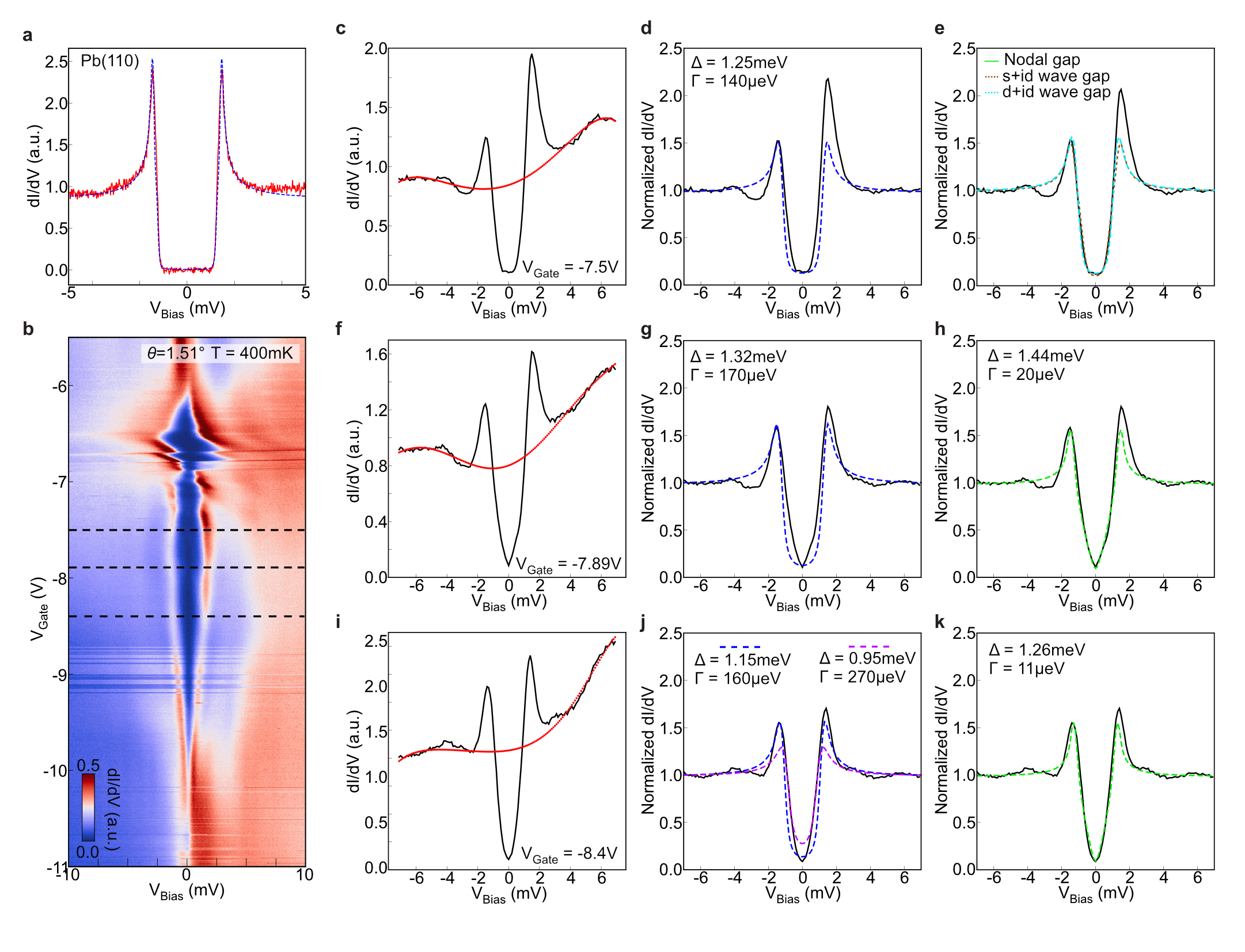}
\end{center}
\caption{{\bf Normalization of tunneling conductance and fitting.} {\bf a}, 
Tunneling conductance measured on Pb (110) surface at $T = 400$~mK showing 
superconducting gap. Blue dashed line is Dynes formula fit with two gaps with following 
parameters, $\mathrm{\Delta_1 = 1.42}$~meV, $\mathrm{\Delta_2 = 1.26}$~meV, 
$\mathrm{\Gamma = 10}$~$\mu$eV, $\mathrm{T = 400}$~mK used to obtain the base 
temperature. {\bf b}, Same data as \prettyref{fig: fig3}a showing larger 
$V_{\rm Bias}$ range. Black dashed lines mark gate 
voltages $V_{\rm Gate} = -7.5, -7.89, -8.4$~V with the corresponding 
line traces shown in subsequent panels. {\bf c}, Line cut in the U-shaped regime 
($V_{\rm Gate} = -7.5$~V). Red dotted line is polynomial fitting curve 
obtained as described in section \ref{methods: fitting}. {\bf d}, Normalized data obtained by 
dividing the raw data (black line in {\bf c}) by polynomial fit (red line in {\bf c}). 
Blue line is Dynes formula fit with isotropic gap. {\bf e}, Same data as {\bf d} with 
Dynes formula fits using different types of the pairing gap symmetry: a nodal gap with 
$\Delta_d = 1.40$~meV (green); $s+id$ pairing gap with 
$\Delta_s = 0.72$~meV, $\Delta_d = 1.22$~meV (brown); 
$d+id$ pairing gap with $\mathrm{\Delta_{d1} = 1.00}$~meV, 
$\mathrm{\Delta_{d2} = 1.30}$~meV (cyan). {\bf f}, in the V-shaped regime ($V_{\rm Gate} = -7.89$~V). 
{\bf g}, Normalized data from {\bf f} and Dynes formula fit using an isotropic gap (blue). 
{\bf h}, Normalized data from {\bf f} with Dynes formula fits using a nodal gap with 
$\Delta = 1.44~\text{meV}$ (green). {\bf i}, Another linecut in the V-shaped regime 
($V_{\rm Gate} = -8.4~\text{V}$). {\bf j}, Normalized data from {\bf i} and Dynes formula fit using an 
isotropic gap (blue, purple). {\bf k} Normalized data from {\bf i} and Dynes formula fits green line is nodal gap with $\Delta = 1.26~\text{meV}$.
}
\label{exfig: fig7}
\end{figure}
\clearpage

\begin{figure}[p]
\begin{center}
    \includegraphics[width=\linewidth]{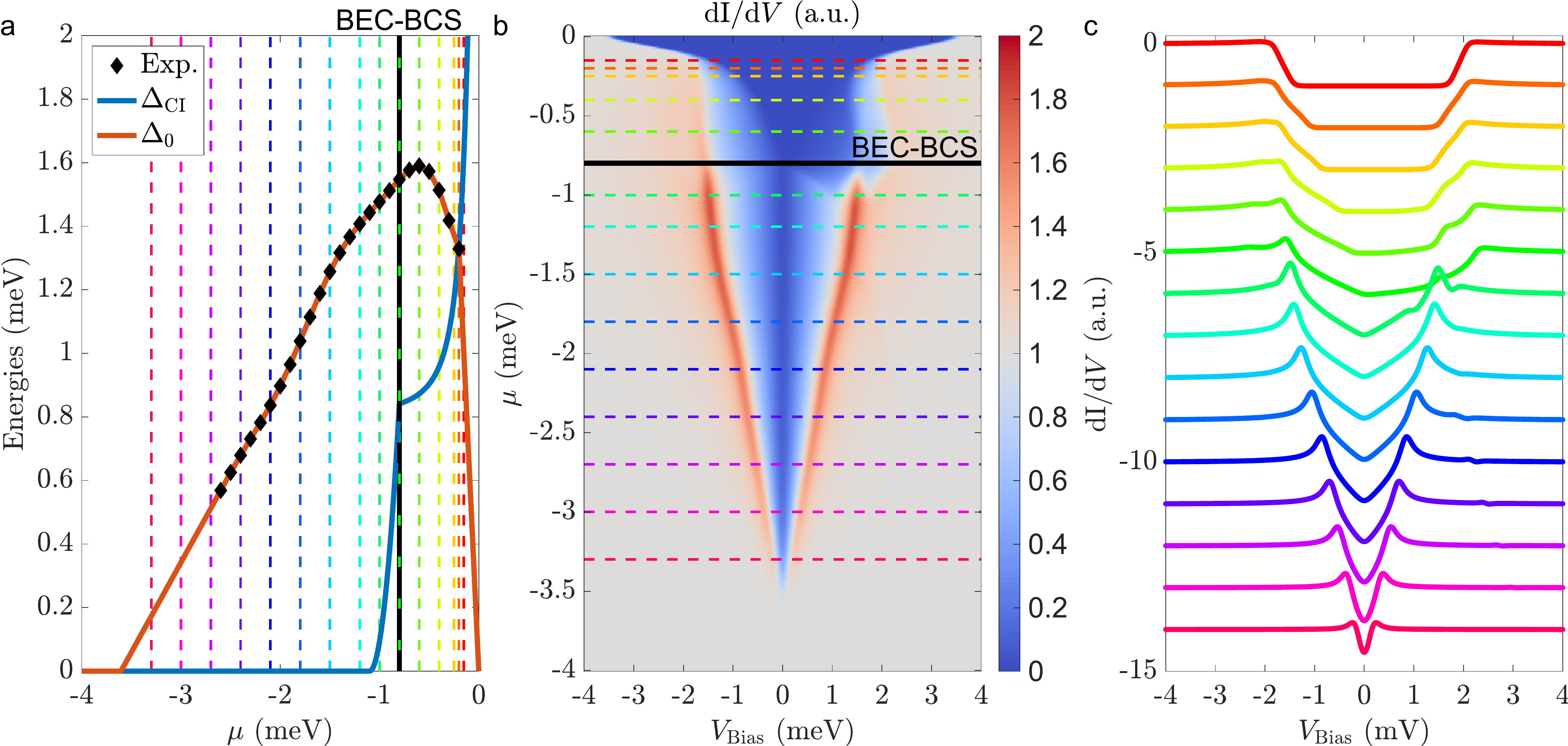}
\end{center}
\caption{{\bf Simulated tunneling conductance across the BEC-BCS transition.} {\bf a}, 
Chemical potential dependence of $\Delta_0$ and $\Delta_{\rm CI}$ used in simulations.  
Black data points represent coherence-peak locations crudely extracted from experiment, as 
detailed in the text.  
{\bf b},{\bf c}, Color map and linecuts of differential conductance $dI/dV$ as a function 
of $\mu$. Here and in \prettyref{fig: fig3}k,j, we set $T=0.05$~meV and employed a nodal 
$d$-wave gap with $1/2 m\ell^2 =6.25~\mu$eV. The BEC-BCS transition manifests as a clear 
evolution from U- to V-shaped spectra as observed experimentally.  We nevertheless stress, as 
in the text, that panels {\bf b},{\bf c} do not correspond directly to Fig.~\ref{fig: fig3}a 
due in part to the nontrivial relation between chemical potential $\mu$ and filling that has 
not been incorporated.} 
\label{fig:theory_figure_methods}
\end{figure}
\clearpage

\begin{figure}[p]
\begin{center}
    \includegraphics[width=14cm]{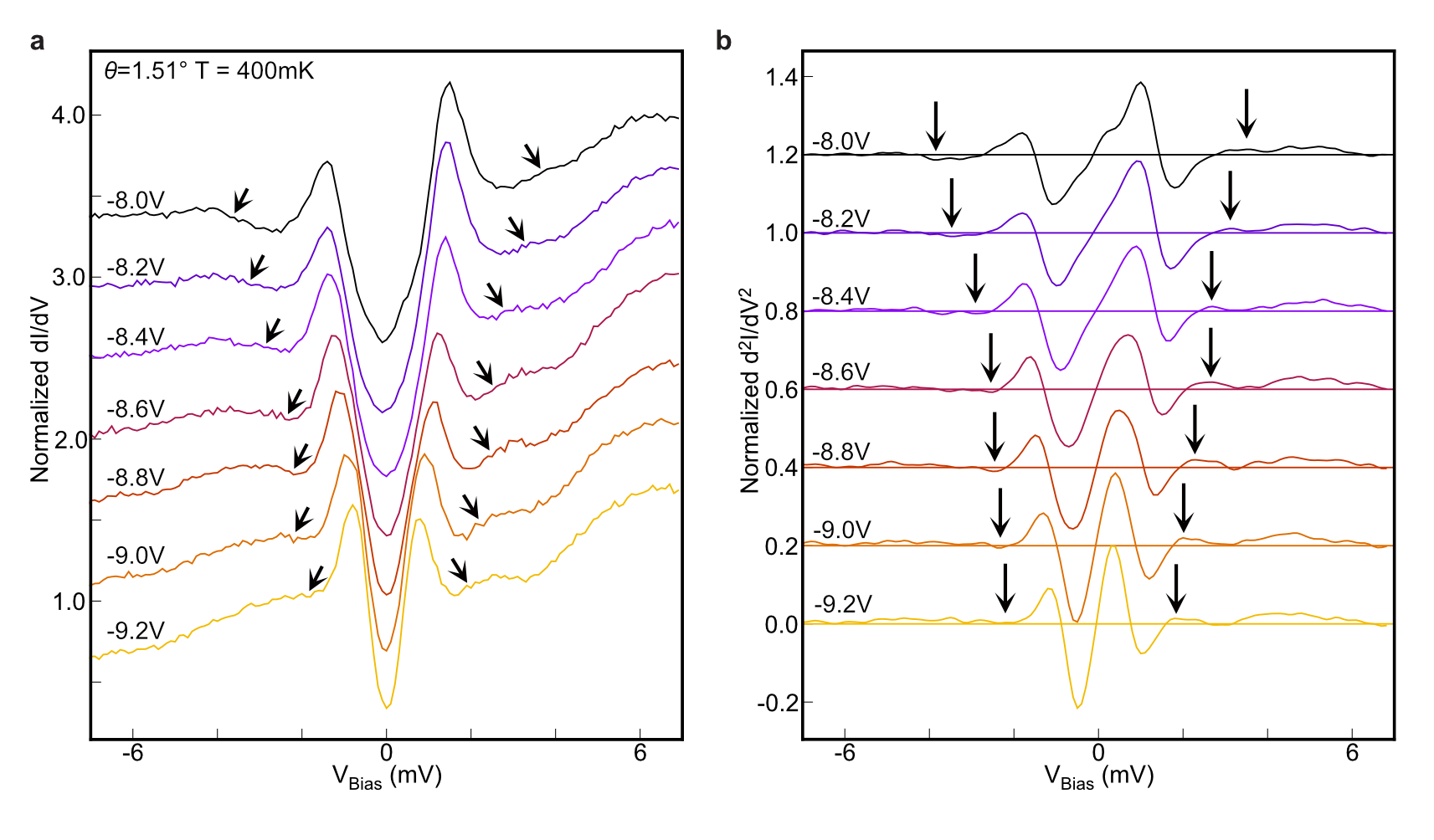}
\end{center}
\caption{{\bf Peak-dip-hump analysis from $d^{2}I/dV^{2}$ local
minima/maxima.} {\bf a}, Hole-side superconducting gap spectrum measured at
various $V_{\rm Gate}$ ranging from $-8.0~\text{V}$ to $-9.2~\text{V}$ at
$\theta = 1.51\degree$ region which is same dataset as
\prettyref{fig: fig4}a. {\bf b}, $d^{2}I/dV^{2}$ as a function of 
$V_{\rm Bias}$ by taking the first derivative of the ({\bf a}) and apply
Gaussian filtering to make the trend clear. The horizontal lines of the same
color indicate the $d^{2}I/dV^{2} = 0$ for each $V_{\rm Gate}$.}
\label{exfig: fig9}
\end{figure}
\clearpage

\begin{figure}[p]
\begin{center}
    \includegraphics[width=14cm]{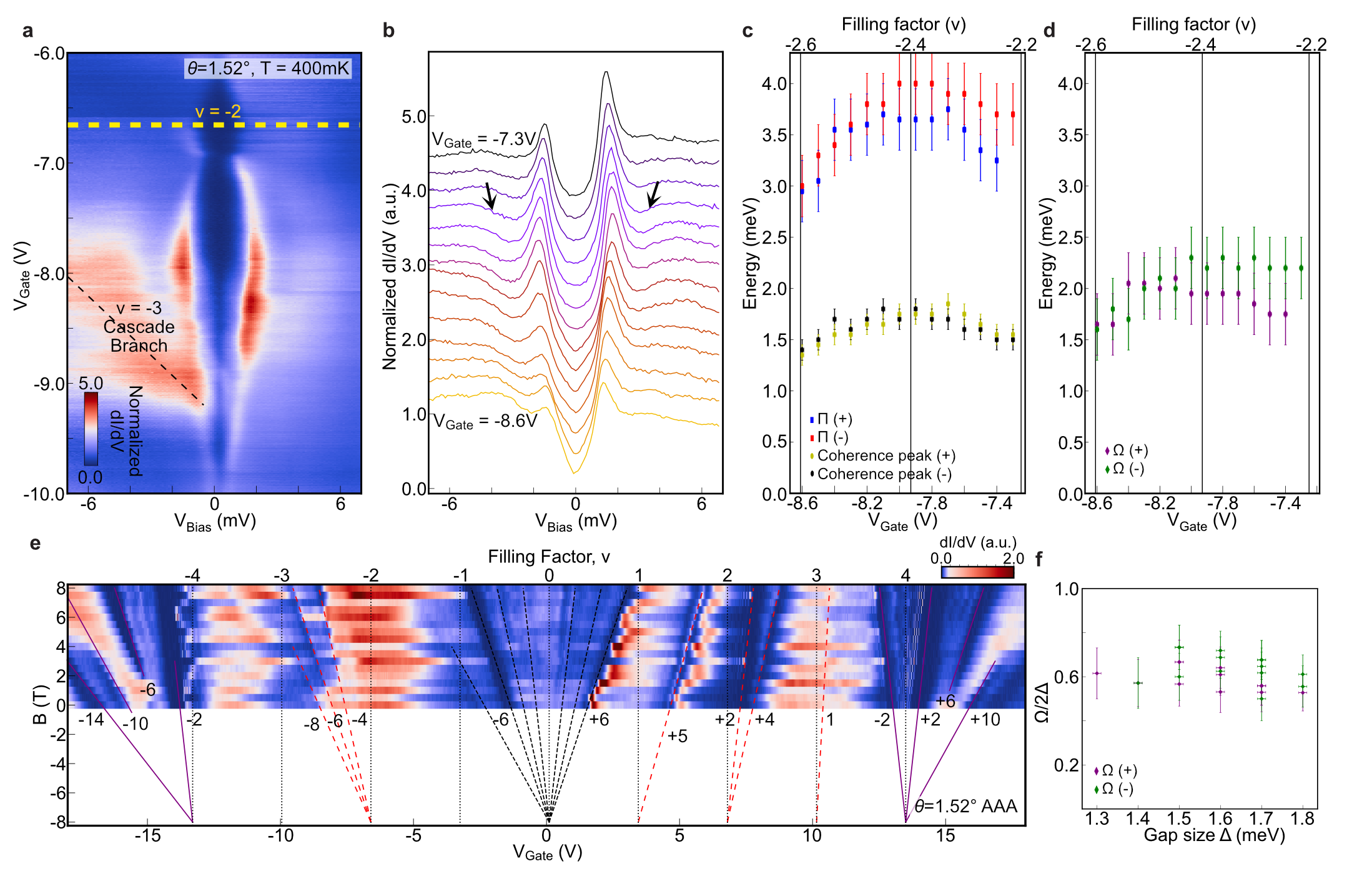}
\end{center}
\caption{{\bf Dip-hump structures observed at different magic-angle area}
{\bf a}, Gate spectroscopy measured at $\theta = 1.51\degree$. 
{\bf b}, Normalized point spectra at range of $V_{\rm Gate}$ from
$-8.6~\text{V}$ to $-7.3~\text{V}$. {\bf c}, Extracted position 
of the dip-hump and a coherence peak versus $V_{\rm Gate}$ for 
$V_{\rm Bias}>0$ (blue and yellow, respectively) and for 
$V_{\rm Bias}<0$ (red and black, respectively). {\bf d}, Energy of 
the bosonic mode versus $V_{\rm Gate}$, obtained by subtracting 
the corresponding energies of the dip-hump feature and the coherence 
peak for $V_{\rm Bias}>0$ (purple) and $V_{\rm Bias}<0$ (green). 
{\bf e}, LDOS Landau fan diagram measured at the same area as {\bf a} 
on AAA region. Black lines indicate the gap between LLs emanating from CNP. 
Red dashed lines indicate gaps between LLs emanating from integer filling 
$\nu\neq0$ of the flat bands. {\bf f}, $\Omega / 2\Delta$ versus 
$\Delta$ obtained from {\bf c},{\bf d}. In this particular area the dip-hump 
structure could be resolved mostly in U-shaped regime.}
\label{exfig: fig10}
\end{figure}
\clearpage

\end{document}